\documentclass[
aps,
pra,
showpacs
amsmath,
amssymb,
preprint,
floatfix,
lengthcheck,
longbibliography
]{revtex4-2}

\usepackage{graphicx}
\usepackage{mathtools}
\usepackage{braket}
\usepackage{hyperref}
\usepackage{bbold}
\usepackage{calrsfs}
\usepackage{dsfont}
\usepackage{cancel}
\usepackage{mathrsfs}
\usepackage[normalem]{ulem}
\usepackage{soul}
\setstcolor{red}% Strike out color

\usepackage{color}

\begin{document}

%-----------------------------------------------------------------
%                      Title and Authors                       
%-----------------------------------------------------------------

\title{Cavity-Mediated Molecular Entanglement and \\ 
Generation of Non-Classical States of Light}

\author{Davis M. Welakuh}
\email{daviswelakuh@g.ucla.edu}
\affiliation{Division of Physical Sciences, College of Letters and Science, University of California, Los Angeles, CA, USA}

\author{Spyros Tserkis}
%\email{spyrostserkis@gmail.com}
\affiliation{Division of Physical Sciences, College of Letters and Science, University of California, Los Angeles, CA, USA}

\author{Scott E. Smart}
\affiliation{Division of Physical Sciences, College of Letters and Science, University of California, Los Angeles, CA, USA}

\author{Prineha Narang}
\email{prineha@ucla.edu}
\affiliation{Division of Physical Sciences, College of Letters and Science, University of California, Los Angeles, CA, USA}

%-----------------------------------------------------------------
%                      Abstract                       
%-----------------------------------------------------------------

\begin{abstract}
The generation and control of entanglement in a quantum mechanical system is a critical element of nearly all quantum applications. Molecular systems are a promising candidate, with numerous degrees of freedom able to be targeted. However, knowledge of inter-system entanglement mechanisms in such systems is limited. In this work, we demonstrate the generation of entanglement between vibrational degrees of freedom in molecules via strong coupling to a cavity mode  driven by a weak coherent field. In a bi-molecular system, we show entanglement can not only be generated between the cavity and molecular system, but also between molecules. This process also results in the generation of non-classical states of light, providing potential pathways for harnessing entanglement in molecular systems.

%Entanglement is vital in many quantum applications ranging from quantum computing to communication and sensing. Molecules have been proposed as a suitable platform for quantum technology applications due to their rich internal structure, comprising of electronic, vibrational, and  rotational degrees of freedom. Entangling those degrees of freedom offers unique opportunities in quantum technology, especially in the construction of quantum memories. In this work, we demonstrate how to induce entanglement between vibrational degrees of molecules by coupling a pair of molecules strongly to a cavity mode driven by a weak coherent field. First, we show that the coupled subsystems (molecules and photon mode) are entangled, and that the entanglement varies with the light-matter coupling strength. Also, we demonstrate a cavity-induced molecule-molecule entanglement. Finally, we demonstrate that non-classical quantum states of light, i.e., states with negative Wigner function, can be generated in the strongly coupled light-matter system. Our results provide useful insights on the emergence on molecular entanglement for strongly coupled light-matter systems and potential for generating non-classical states of light for applications in quantum technologies.
\end{abstract}

\maketitle

%-----------------------------------------------------------------
%                      Main Text                      
%-----------------------------------------------------------------

\section{Introduction}
\label{sec:introduction}

%A vital resource for numerous quantum applications is the property of entanglement which allows two or more quantum states to become correlated in a non-classical way~\cite{Plenio_Virmani_B_14}. While quantum entanglement has been observed and studied extensively with individual atoms and photons~\cite{mcConnell2015,hacker2019,lin2020}, recent advances in experimental techniques and theoretical understanding have enabled investigations into entanglement with larger systems, including molecules~\cite{deMille2002,carr2009,blackmore2019,ding2023}. This is due to their unmatched flexibility in terms of material composition, complex internal structure, fine structural tuning, integration into photonic structures, and long-lived interaction states that molecules possess. Thus, being able to control molecules at a quantum level, would give access to further degrees of freedom such as the vibrational or rotational degrees, and entangling those degrees of freedom offers unique opportunities in quantum information processing, especially in the construction of quantum memories~\cite{lenz2021,serrano2022,albert2020}. 
 
 Quantum entanglement is a vital resource for numerous quantum applications, which allows two or more quantum states to become correlated in a non-classical way~\cite{Plenio_Virmani_B_14}. While quantum entanglement has been observed and studied extensively with individual atoms and photons~\cite{mcConnell2015,hacker2019,lin2020}, recent advances in experimental techniques and theoretical understanding have enabled investigations into entanglement with larger systems, including molecules~\cite{deMille2002,carr2009,blackmore2019,ding2023}. Molecular systems have unmatched flexibility in terms of material composition, complex internal structure, fine structural tuning, integration into photonic structures, and long-lived interaction states. Thus, being able to control and entangling various molecular degrees of freedom offers unique opportunities in quantum information processing and quantum memories~\cite{lenz2021,serrano2022,albert2020}. 

Over the past few years, the strong coupling of quantum light and matter via infrared or optical cavities has experienced a surge of interest in chemistry and material science as a new approach for controlling or modifying chemical reactivity and physical properties. Seminal experimental work has demonstrated the possibility to control photochemical reactions~\cite{hutchison2012,kowalewski2016} and energy transfer~\cite{coles2014,zhong2016}, enhancement of harmonic generation from polaritonic states~\cite{chervy2016,barachati2018,welakuh2022b,welakuh2022c}, modification of ground-state chemical reactions via vibrational strong coupling~\cite{thompson2006,thomas2019}, or cavity-control of condensed matter properties~\cite{latini2021,liu2015}. In addition, for vibrational strong-coupling (VSC), it was observed that coupling to specific vibrational excitations can inhibit~\cite{thomas2016}, steer~\cite{thomas2019} and enhance~\cite{lather2020} molecular reaction rates. Recent theoretical investigations have focused on how light and matter become entangled and what happens to the entanglement with increasing temperature. Examples includepolaritonic systems under VSC in thermodynamic equilibrium~\cite{sidler2023} even with disorder~\cite{wellnitz2022}, cavity–catalyzed hydrogen transfer~\cite{fischer2023}, and others. Importantly, these works mainly focus on entanglement between light and matter (molecule) and not the entanglement between the matter sector due to the quantized field for the case of a molecular ensemble. 
%Also, the important aspect of decoherence that presents a realistic description of the interacting strongly coupled system is not usually considered in these investigations, mainly due to large dimensionality from an ab-initio perspective.

In this work, we close this gap by investigating the onset of entanglement between molecules strongly coupled to a cavity and driven externally by a coherent field. We consider the vibrational stretch mode of an O$-$H bond of two identical non-interacting molecules within an infrared cavity in which a mode frequency is resonant to the lowest vibrational transition. We find that the photon mode and molecules become entangled for the entirety of the simulation, and that the entanglement is tunable with respect to the light-matter coupling strength.  In addition, we show a cavity-induced molecule-molecule entanglement via the logarithmic negativity of the molecular subsystem. %By this account, the populated cavity mode induced entanglement between the vibrational stretch mode of the O$-$H bond between the two molecules. 
%On the chemistry side, this can explain energy transfer between vibrational states of molecules~\cite{zhong2017}.
%To account for dissipation and decoherence channels in the coupled system, we explicitly couple to a photonic bath and show that the entanglement induced by the cavity mode is robust to such effects when the vibrational mode of the molecules are strongly coupled to the enhanced cavity mode. In addition, we show that upon increasing the light-matter coupling strength, the degree of entanglement can be increased as the temporal profile of the von Neumann entropy increases with increasing light-matter coupling strengths. This effect has the potential to overcome decoherence in some systems with short coherence times, since features showing strong entanglement can be caused to occur at earlier times by increasing the coupling strength.
Lastly, we investigate the possibility that the initially coherent drive field becomes modified due to its interaction with a strongly coupled cavity-matter system. We show that the initially coherent field becomes non-classical at a later time as negative regions emerge in the Wigner function of the photon mode which indicates the quantum state of the photon field is non-classical. Collectively these results highlight a robustness in the generation and control of entanglement in the coupled system, as well as the complexity which emerges in the molecular and cavity systems. 

%These result highlights that for a strongly coupled molecular system, we can on the one hand modify matter properties and on the other hand generate non-classical states of light which have very promising applications in quantum information processing~\cite{ralph2003}.

\section{Theoretical framework}
\label{sec:general-framework}

For a fundamental quantum-mechanical description of the light-matter interaction, we consider the non-relativistic quantum electrodynamics (QED) Pauli-Fierz Hamiltonian in the long-wavelength limit~\cite{tannoudji1989,spohn2004}. In this approximation the relevant photon modes in the infrared or optical regime have wavelengths that are large compared to the size of a matter subsystem. Within the setting of a nuclear–photon interaction, the length form of the Pauli-Fierz Hamiltonian~\cite{rokaj2017,flick2019} is given by
\begin{align} 
\hat{H} =&\sum\limits_{i=1}^{N}\left(\frac{\hat{\textbf{p}}_{i}^{2}}{2\mu} + V(\hat{\textbf{r}}_i)\right) \nonumber \\ 
&+ \frac{1}{2}\sum_{\alpha=1}^{M}\left[\hat{p}^2_{\alpha}+\omega^2_{\alpha}\left(\hat{q}_{\alpha} \!-\! \frac{\boldsymbol{\lambda}_{\alpha}}{\omega_{\alpha}} \cdot \hat{\textbf{R}} \right)^2\right].\label{eq:el-pt-hamiltonian}
\end{align}
Here, $\mu$ is the reduced molecular mass and we consider a single vibrational degree of freedom of $N$ molecules with vibrational coordinates, $\hat{\textbf{r}}_{i}$, and their conjugate momenta $\hat{\textbf{p}}_{i}$. The quantized electromagnetic field is described by a collection of harmonic oscillators which consist of the displacement coordinate $\hat{q}_{\alpha}$ and canonical momentum operator $\hat{p}_{\alpha}$ with associated mode frequency $\omega_{\alpha}$ for each mode $\alpha$ of an arbitrarily large but finite number of photon modes $M$. The coupling strength between light and matter is given as $\boldsymbol{\lambda}_{\alpha} = \lambda\textbf{e}_{\alpha}$ where  $\textbf{e}_{\alpha}$ is the polarization vector of the photon modes. A variety of potentials $V(\hat{\textbf{r}})$ can be used to describe different physical and chemical effects that occur in the vibrational sector of the coupled system. In this work we consider the vibrational stretch mode of an O$–$H bond and describe the physical setting using a Morse potential
\begin{align} 
 V(\hat{\textbf{r}}) = D_{e}\left( e^{-a\left(\hat{\textbf{r}} - \textbf{r}_{e} \right)} - 1 \right)^{2} , \label{eq:morse-potential}
\end{align}
where $D_{e}$ represents the depth of the potential well, and the parameter $a$  determines the ``width'' of the potential well. The importance of the Morse potential is that it gives a reasonably realistic representation of a bonding potential energy curve, while at the same time being exactly solvable. Next, we define the electron-photon coupling from the bilinear interaction term of Eq.~(\ref{eq:el-pt-hamiltonian}) as 
\begin{align}
g_{\alpha}^{(ij)}  = \lambda\sqrt{\frac{\hbar\omega_{\alpha}}{2}} \langle \varphi_{i}|\textbf{e}_{\alpha}\cdot\hat{\textbf{R}}|\varphi_{j} \rangle \, ,
\end{align}
where $|\varphi_{j} \rangle$ are the vibrational eigenstates and $\langle \varphi_{i}|\textbf{e}_{\alpha}\cdot\hat{\textbf{R}}|\varphi_{j} \rangle$ is the transition dipole moment. We define the dimensionless ratio $\eta = g_{\alpha}/\hbar\omega_{\alpha}$ which characterizes the strength of the light–matter interaction in relation to the matter excitations. Conventionally, the change from the vibrational strong coupling to the vibrational ultra-strong coupling regime is indicated by values of $\eta \geq 0.1$~\cite{kockum2019}.

In the following, we investigate for the strongly coupled light-matter system, how the photon mode induces entanglement between vibrational states of identical molecules. Commonly the vibrational degrees of freedom are mostly considered as decoherence channels for electronic excitations with less focus on their quantum-mechanical nature which has potential applications in quantum technologies. In fact, molecular systems can allow for representations of qubits that are more complex~\cite{wernsdorfer2019,wasielewski2020}, where decoherence sources can be reduced or managed by the molecular composition~\cite{wasielewski2020}. Based on this, it is particularly important to show that strongly coupled vibrational degrees to a cavity field can show robust quantum entanglement even in the presence of decoherence. This work highlights that such systems provides a potential platform for the development and implementation of future quantum technologies.

\section{Molecule-molecule entanglement via strong coupling}
~\label{sec:molecule-entanglement} 

\begin{figure}[t!]
    \includegraphics[width=3.0in,height=3.8in]{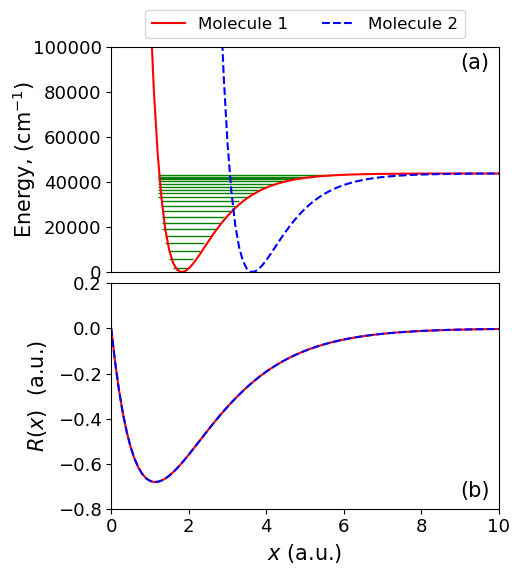}
\caption{(a) The Morse potential for the stretching modes of an O$-$H for two identical molecules including the bound vibrational states which displays the anharmonicity of the potential. The displacement of the Morse potential for molecule 2 is for visual purposes only. Panel (b) shows the corresponding dipole function for the O$-$H stretching mode which is nonlinear for diatomic molecules.}
\label{fig:morse-potential}
\end{figure}

Focusing on quantum states described by Eq.~(\ref{eq:el-pt-hamiltonian}), we want to study how entanglement behaves for certain bipartitions of the system. We consider the vibrational stretching mode of an O$–$H bond for two identical molecules confined within a photonic environment. First, we describe the details of the vibrational degrees of the coupled system. For the stretching mode of the O$-$H bond along one spatial dimension, the values of the parameters of the Morse potential, taken from Refs.~\cite{korolkov1996,paramonov2009}, are $D_{e}=0.1994$ E$_{h}$, $r_{e}=1.821$ a$_{0}$, and $a=1.189$ a$_{0}^{-1}$ where a$_{0}$ represents the Bohr radius and E$_{h}$ the Hartree. To describe the diatomic O$-$H stretching mode, we adopt the Mecke dipole function~\cite{mecke1950} which provides a fairly accurate description of the molecule's dipole moment $\hat{\textbf{R}}(\hat{x})=-\gamma \, \hat{x} e^{-\delta \hat{x}}$ where the parameters are chosen to fit experimental data. The values of the parameters for the O$-$H bond stretch  are $\delta=0.8818$ a$_{0}^{-1}$ and $\gamma=1.634$~\cite{korolkov1996,paramonov2009}. In Fig.~(\ref{fig:morse-potential}), we show the Morse potential that describes the vibrational stretch mode of the O$-$H bond for two identical non-interacting molecules including the bound vibrational states and the nonlinear dipole function. We note that Eq.~(\ref{eq:el-pt-hamiltonian}) does not include the interactions between the molecules. This is because we want to investigate the influence the quantized photon field has in mediating molecule-molecule entanglement. 

When the molecules are confined within an infrared cavity, we choose a cavity mode with frequency that is resonant to the two lowest vibrational states of both molecules indicated in Fig.~(\ref{fig:morse-potential}a) with transition frequency $\hbar\omega_{01}=3783.267$ cm$^{-1}$. The corresponding dipole moment for this transition is $\langle \varphi_{i}|\textbf{e}_{\alpha} \!\cdot\! \hat{\textbf{R}}|\varphi_{j} \rangle \!=\! 0.025$ a$_{0}$ and for a coupling parameter $\lambda=0.01$ we obtain the dimensionless value $\eta = 0.001$ which is within the strong coupling regime. An increasing value of $\eta$ {(for $\eta<0.1$) indicates that we exploring effects in the} strong coupling regime. To simulate the dynamics of such a coupled system, we explicitly propagate the time-dependent Schr\"{o}dinger equation $i\hbar \frac{\partial}{\partial t} |\Psi (t)\rangle = \hat{H}|\Psi (t)\rangle$ with the Hamiltonian of Eq.~(\ref{eq:el-pt-hamiltonian}). As initial state, we consider a factorizable product state of the form $|\Psi_{\text{in}} (0)\rangle = |\varphi_{0}\rangle_{1}|\varphi_{0}\rangle_{2}|\beta\rangle$, where $|\varphi_{0}\rangle_{1}, \; |\varphi_{0}\rangle_{2}$ are the vibrational ground-states of molecule 1 and molecule 2,  $|\beta\rangle= e^{-|\beta|^{2}/2} \sum_{n=0}^{\infty}  \left(\beta^{n}/\sqrt{n!}\right)|n\rangle$ is a coherent state with amplitude $\beta$ where $|n\rangle$ are the Fock states of the photon mode. The coherent state field that populates the cavity mode with photons is used to induce dynamics in the coupled light-matter system. We choose {as a starting point} the amplitude $\beta=2$, so that at the initial time the infrared cavity is populated with on average $n=|\beta|^{2}=4$ photons. 
%In the case were we include coupling to a photonic bath to account for decoherence in the coupled system, the initial state is a product state of the form $|\Psi_{\text{in}}' (0)\rangle = |\varphi_{0}\rangle_{1}|\varphi_{0}\rangle_{2}|\beta\rangle|0\rangle_{2}|0\rangle_{3}...|0\rangle_{51}$, where $|0\rangle_{2}|0\rangle_{3}...|0\rangle_{51}$ are the vacuum states of the 50 photon modes that are included. The bath modes cover the energy range $219.475$ to $8778.985$ cm$^{-1}$ with equidistant mode spacing $\Delta\omega_{\alpha}=174.685$ cm$^{-1}$ and we choose a coupling to the bath modes with value $\lambda'=0.001$.

\begin{figure}[t!]
    \includegraphics[width=0.8\columnwidth]{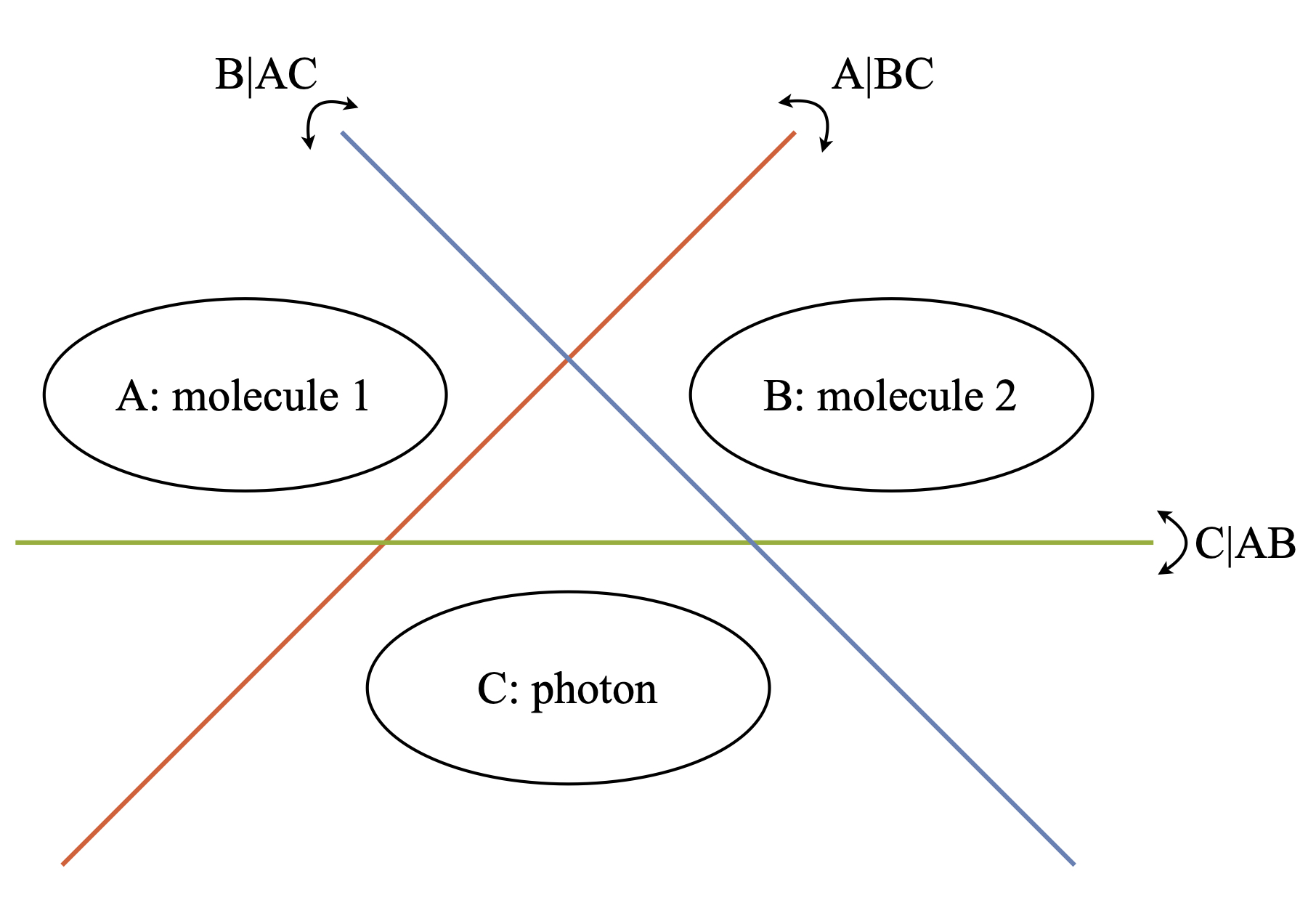}
\caption{Bipartitions of the constituents of the coupled light-matter system. Partitions $\text{A}|\text{BC}$ (red line) and $\text{B}|\text{AC}$ divide molecule 1 from the combination of molecule 2 and the cavity mode and molecule 2 from the combination of molecule 1 and the cavity mode, respectively. Analogously, partition $\text{C}|\text{AB}$ (green line) divides the cavity mode from both molecules 1 and 2. 
}
\label{fig:bipartitions}
\end{figure}

There are several different measures that can be used to characterize the entanglement that arises in the dynamics of the coupled system~\cite{hammerer2010}. In this work, we analyze entanglement between the photonic and nuclear degrees of freedom  by computing two measures depending on the purity of the state. In particular, for pure quantum states we employ the entropy of entanglement~\cite{Bennett_etal_PRA_96} and for mixed states the logarithmic negativity~\cite{plenio2005}. Considering first the entire system, i.e., pure quantum state, we have three constituents, namely: A: molecule 1, B: molecule 2, and C: photonic mode, which give rise to three biparititions of the system: $\text{A}|\text{BC}$, $\text{A}|\text{BC}$, and $\text{A}|\text{BC}$, depicted in Fig. (\ref{fig:bipartitions}). Thus we can quantify the entanglement of all three bipartitions using the entropy of entanglement, defined as
\begin{equation}
   E(\rho) :=  S(\rho_i) ,
\end{equation}
where $\rho_i$ is the reduced density matrix of either bipartition and $S(\rho) := - \textrm{Tr} ( \rho \log_2 \rho)$ the von Neumann entropy. 
%{\color{blue}ST: I moved that information on the next page} \st{This is illustrated in Fig.~(2) where we indicate how we compute the entanglement of all three bipartitions. That is, the partition $\text{A}|\text{BC}$ (red line) indicates the entanglement between molecule 1 and a combination of molecule 2 and the cavity mode. The partition $\text{C}|\text{AB}$ (green line) represents the entanglement between the cavity mode and both molecules 1 and 2 while partition $\text{B}|\text{AC}$ (blue line) indicates the entanglement between molecule 2 and a combination of molecule 1 and the cavity mode}

To investigate the molecule-molecule entanglement of vibrational degrees induced by the driven cavity mode, we rely on {a different measure, due to the fact that the bipartition we are interested in constitutes a mixed quantum state, where entropy of entanglement is not a valid measure any more. We employ instead} the logarithmic negativity which is defined as~\cite{plenio2005}
\begin{align} \label{eq:logarithmic-negativity} 
E_N (\rho) := \text{log}_{2} \| \rho^{\Gamma} \|_1 \, , 
\end{align}
where $\rho^{\Gamma}$ denotes the partial transpose of the density matrix $\rho$ in respect to one of its two partitions and $\| \cdot \|_1$ the trace norm.

\begin{figure}[t!]
    \includegraphics[width=3.0in,height=2.5in]{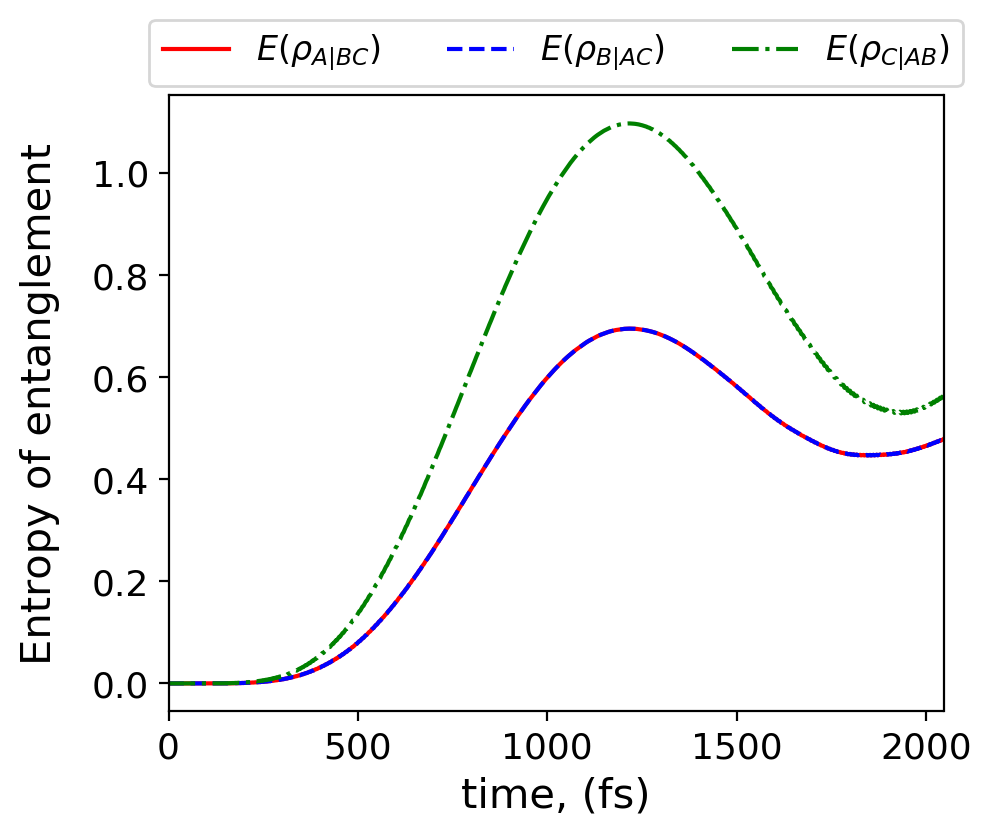}
\caption{Entropy of entanglement $E(\rho)$ between the coupled subsystems in real-time. {The bipartition between the photon and the combination of the two molecules is more entangled when compared to the other two bipartitions.} The dimensionless light-matter coupling here is $\eta=0.001$.}
\label{fig:entropy-purity}
\end{figure}

We now present the results for the investigation of entanglement between the molecules and cavity mode in the strongly coupled system. In Fig.~(\ref{fig:entropy-purity}), we show the real-time entanglement dynamics among the three bipartitions of the system. We denote by $E(\rho_{\text{A}|\text{BC}})$ (red line) the entanglement between molecule 1 and a combination of molecule 2 and the cavity mode, by $E(\rho_{\text{B}|\text{AC}})$ (blue line) the entanglement between molecule 2 and a combination of molecule 1 and the cavity mode, and by $E(\rho_{\text{C}|\text{AB}})$ (green line) the entanglement between the cavity mode and both molecules 1 and 2. We find as expected that the  entropy of entanglement for the three bipartitions are zero at the initial time indicating that the subsystems are decoupled. As time progress, the entanglement deviates from zero and the bipartitions have their respective maximum entanglement at around 1250~fs before slightly losing features of entanglement as time progresses. The populated photon mode is shown to be strongly entangled with the (combination of) molecules, i.e., with both of their vibrational modes, when compared to the vibrational degrees which show an equal entanglement dynamics for the entire simulation. The reason for this is that the identical molecules with the same vibrational transitions of the O$-$H stretch bond is resonantly coupled to the cavity mode with equal probability to coherently exchange excitation with the photon field. {We note that the entropy of the photon mode being non-zero beyond the initial time indicates that the field is in an entangled state with both of the vibrational modes of the two molecules.} {Entanglement on the other two bipartitions implies that} the vibrational degrees of molecule 1 (2) is in an entangled state with both the photon mode and molecule 2 (1). These results show the onset and build up of entanglement in an initially decoupled system due to strong coupling with a photon mode that coherently exchange excitations as a result of resonant light-matter coupling.

\begin{figure}[t!]
    \includegraphics[width=3.0in,height=3.8in]{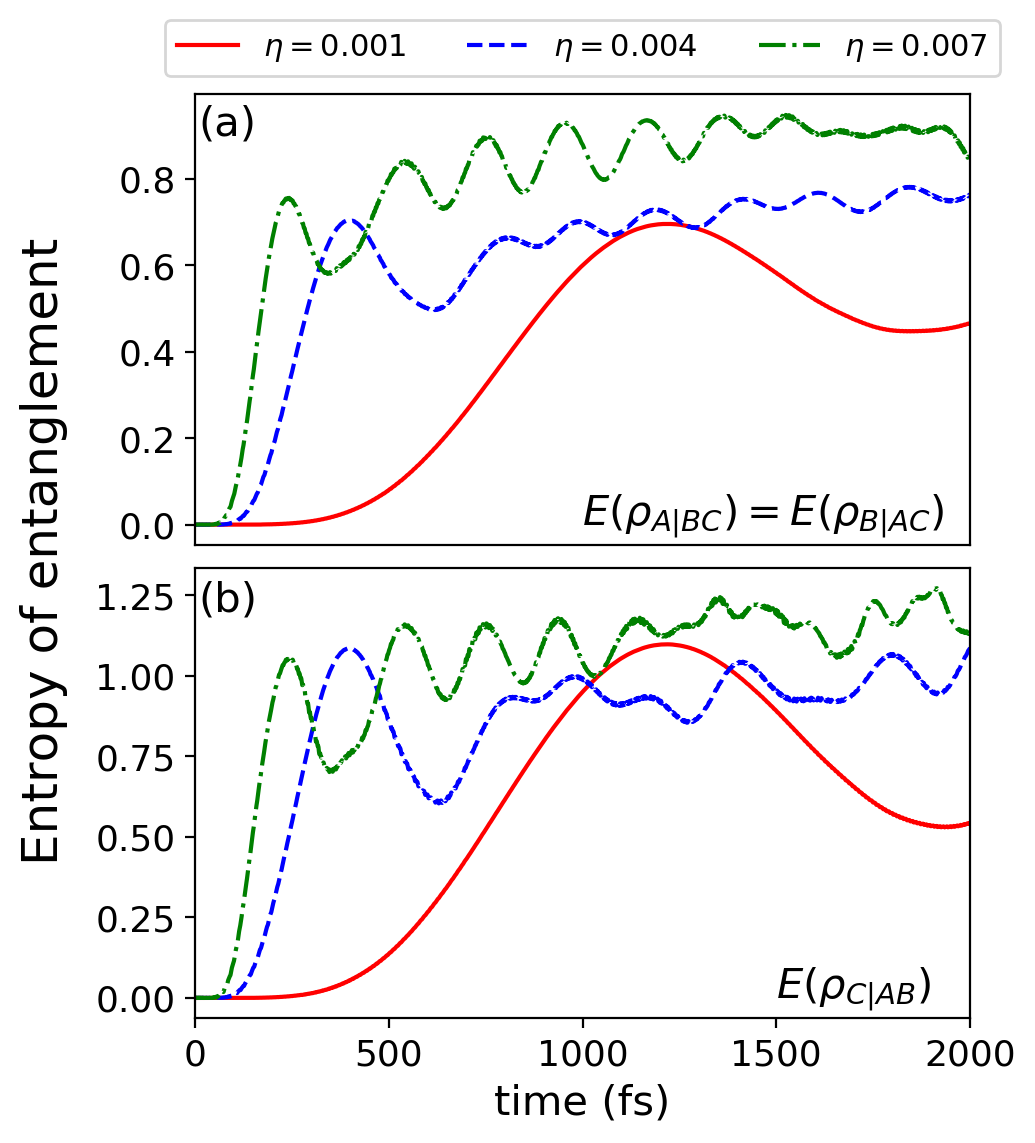}
\caption{(a) The real-time entropy of entanglement between {one molecule and the rest of the system} showing how the degree of entanglement changes with an increasing light-matter coupling parameter $\eta$. Panel (b) shows the entropy of {entanglement between the photon mode and the two molecules} for increasing $\eta$. The degree of entanglement increases and the maximum occurs at earlier times.}
\label{fig:entropy-mols-ptn}
\end{figure}

Next, we investigate the role an increasing light-matter coupling strength has on the entanglement of the coupled subsystems. To do this, we enhance the interaction between the coupled subsystems by increasing the coupling $\lambda$ which intends increases the dimensionless parameter $\eta$ used to quantify the strength of the light-matter interaction. In Fig.~(\ref{fig:entropy-mols-ptn}), we find that increasing the light-matter coupling strength increases the degree of entanglement among the constituents of the strongly coupled light-matter system. Interestingly, we also find that the time at which the molecules and photon mode are strongly entangled can be made to occur at earlier times by increasing the light-matter coupling strength. This particular feature has the potential to provide longer coherence times with strong entanglement features occuring before decoherence sets in at later times. Exploring this feature is particularly relevant for potential quantum-technological applications with molecular qubits~\cite{wasielewski2020}. 

\begin{figure}
    \includegraphics[width=3.0in,height=2.5in]{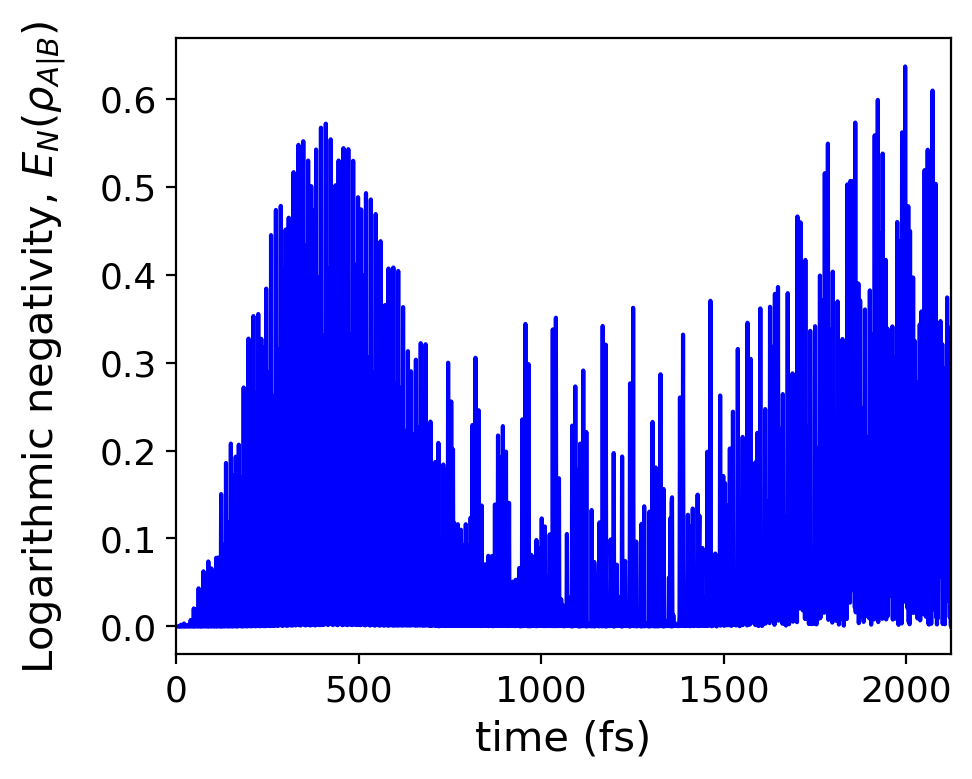}
\caption{Entanglement between the vibrational degrees of the molecular systems induced by the excited cavity mode as measured by the logarithmic negativity. The molecular systems show entanglement after the initial time since $E(\rho)>0$ demonstrating molecule-molecule entanglement induced by the cavity mode.}
\label{fig:entropy-purity-without-bath}
\end{figure}

Finally, we consider the case of the molecule-molecule entanglement of vibrational degrees induced by a cavity mode populated by a coherent state field. Here, tracing out the photon degrees gives information on the role the populated cavity mode contributed in entangling the vibrational stretch mode of two O$-$H bonds. For this investigation, we compute the logarithmic negativity as defined in Eq.~(\ref{eq:logarithmic-negativity}) and show in Fig.~(\ref{fig:entropy-purity-without-bath}) the temporal profile in real-time. At the initial time, the molecules are not entangled since $E_N (\rho)$ is zero which is due to the choice of the initial state being a product state. As time progresses, the profile of the logarithmic negativity is non-zero throughout the entire simulation demonstrating that the populated cavity mode induces molecule-molecule entanglement between the vibrational stretch mode of the O$-$H bonds of the two molecules. The onset of entanglement between the molecules due to the driven cavity mode might also be used to explain the energy transfer between spatially separated molecules~\cite{zhong2017}. 

\begin{figure*}[bth]
\includegraphics[width=5.5in,height=4.5in]{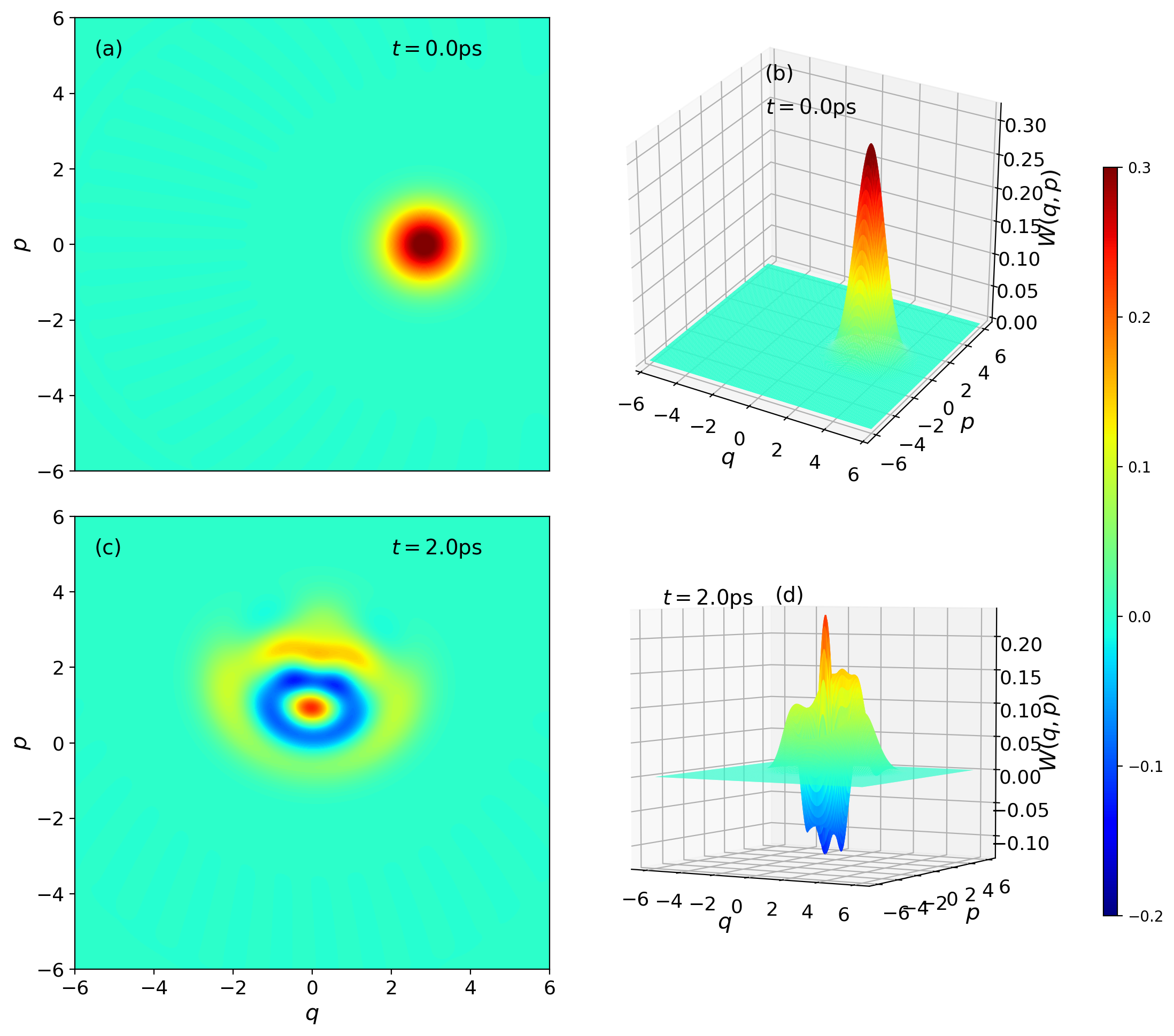}
\caption{(a) A two-dimensional (2D) representation of the Wigner function of the coherent state field at the initial time and panel (b) shows the corresponding three-dimensional (3D) representation. Panels (c) and (d) show the corresponding 2D and 3D plots at time $t=2$~ps with non-classical features as negative regions appear in the Wigner function. }
\label{fig:wigner-function}
\end{figure*}

In the results presented, we have not included natural decoherence or dissipation channels like coupling to the environment. For strongly coupled light-matter systems where the cavity has an influence on coupling to the environment, this can be modeled by coupling the internal system to a photonic bath. By doing this, we expect that the features of entanglement presented above will not deviate strongly but there will be a decrease in the degree of entanglement depending on how strongly the photonic bath modes are coupled to the internal system. Usually, in the strong coupling regime the coupling between the internal structure and photonic bath is weaker than the light-matter coupling strength which allows the physical effects under study to remain qualitatively unchanged as in Ref.~\cite{welakuh2021}. From this, we expect the  molecule-molecule entanglement of vibrational degrees induced by the populated cavity mode to be qualitatively similar when coupled to a photonic bath.

%In the results discussed so far, we did not include the naturally occurring decoherence channels (i.e. coupling to environment) that exist in such strongly coupled systems. In order to include such contributions, we include a coupling to a photonic bath~\cite{welakuh2021,wang2021} sampled here with 50 photon modes as discussed above. In Fig.~(\ref{fig:entropy-purity-bath}), we show a comparison between the von Neumann entropy for the case where we do not couple to a bath and the case where coupling to the photonic bath is included. Qualitatively, both simulations show the same behavior, with the main difference that the degree of entanglement is reduced when the bath modes are included and the time at which the entanglement is maximum appears slightly later. If we simulate for a longer time, we expect the differences to become more pronounced and provided we do not simulate for a very long time such that we hit the unphysical revival time due to having only finitely many modes~\cite{bocchieri1957}. The results of Fig.~(\ref{fig:entropy-purity-bath}) highlight that the entanglement induced in between the nuclear degrees and photonic mode is robust which is particularly important for quantum technologies.

\section{Generation of non-classical states of light}
~\label{subsec:non-classical-states}

In the developing fields of polaritonic chemistry and materials in cavities~\cite{kockum2019,ruggenthaler2017b,flick2018a}, the main focus is on how the strongly coupled photons influence matter properties, and less focus is usually given on how the photon field gets modified due to its interaction with matter. Within our study of entanglement induced by strong light-matter coupling, we investigate properties of the photon field as it interacts strongly with matter. The main goal is to investigate whether the induced entanglement leads to the possibility of generating non-classical states of light when the populated cavity mode interacts strongly with the molecular systems. Specifically, we investigate whether the initial coherent state field used as external drive remains coherent throughout the entire interaction with the nuclear degrees. The Wigner function provides a natural representation of quantum states of the cavity field as it is a useful criteria to determine the nonclassical feature of the field. In our study, we compute the Wigner function of the photonic mode which is defined as~\cite{wigner1932}
\begin{align}
W(q, p) =\frac{1}{2\pi\hbar} \int_{-\infty}^{\infty} dq' e^{\frac{ipq'}{\hbar}} \langle q - \frac{q'}{2}|\hat{\rho}|q + \frac{q'}{2}\rangle \, , \label{eq:wigner-function} 
\end{align}
where $\hat{\rho}$ is the reduced density matrix of the photonic sector of the strongly coupled light-matter system. The Wigner function can be negative {or singular} in some regions of phase space {which} is an indication of a nonclassical {behaviour}. It is important to note that the negativity of the Wigner function is a sufficient but not necessary condition for nonclassicality of a quantum state~\cite{walls1994}. 

In Fig.~(\ref{fig:wigner-function}), we show the Wigner function at two different times during the simulation. At the initial time, the cavity mode is populated with a coherent state as represented in the two-dimensional (in Fig.~(\ref{fig:wigner-function}a)) and three-dimensional (in Fig.~(\ref{fig:wigner-function}b)) projection of the Wigner function, where a Gaussian distribution is depicted for the coherent state. At a later time when $t=2.0$~ps, we find that the initially Gaussian distribution representing the coherent state in phase space now has regions with negative values of the Wigner function (see Fig.~(\ref{fig:wigner-function}c,d)). The negative regions of the Wigner function demonstrates that the state of the photon field is non-classical at this point in time. We note that the non-classical feature (negative values of the Wigner function) arises at time $t=1.35$~ps and is long-lived for the entire simulation. This results highlights that non-classical states of the cavity field are generated in the case where it interacts strongly with the vibrational degrees of molecules. In addition, this results suggests that the focus of polaritonic chemistry and materials in cavities should also consider investigating how the matter system can modified the photon field for the case when an external driving is included.

\section{Conclusions and Outlook}
\label{sec:conclusion-outlook}

In this work we investigated the coherent femtosecond dynamics of vibrational degrees of two molecules strongly coupled to an excited cavity mode in a coherent state. Specifically, we analyzed the onset of cavity-induced molecule-molecule entanglement for a coherently driven coupled light-matter system. First, we showed that when the excited cavity mode is resonantly coupled to the lowest vibrational transition of the O$-$H stretch bond of the two molecules, the {entanglement} of the different {bipartitions} is non-zero for the entire simulation. This demonstrates that {all the bipartitions} of the coupled system become entangled with each other after the excitation by the coherent state field. Secondly, we investigated the role of strong light-matter coupling on the entanglement between the molecules and cavity mode where we found that entanglement can be increased by increasing the light-matter coupling strength and with this, features of entanglement can be pushed to occur at earlier times. This is particularly important since for systems with short coherence times, strong light-matter coupling can cause desired physical effects to occur at earlier times when the system is still coherent by increasing the light-matter coupling strength. Furthermore, we investigated the molecule-molecule entanglement induced by the cavity mode where we showed that the molecules become entangled since the logarithmic negativity is non-zero for the entire simulation. This result highlights the role the cavity mode plays in entangling the vibrational degrees of the molecules. Lastly, we investigated the possibility of generating non-classical states of light via strong light-matter interaction. For the same system and setup, we computed the Wigner function of the excited cavity mode in a coherent state which at the initial time is a Gaussian distribution in the phase space picture. We also showed that the phase space picture has regions with negative values of the Wigner function which indicates the generation of non-classical states of light of the coupled light-matter system. 

The generation and control of genuine quantum mechanical entanglement between molecules as a result of the light matter interaction yields an important tool for the construction of novel quantum materials. These result presents a new perspective for polaritonic chemistry and materials in cavities towards investigating how matter systems can modify the photon field in a strongly coupled light-matter system with an external excitation.

%-----------------------------------------------------------------
%				               Acknowledgement 
%-----------------------------------------------------------------

\section*{Acknowledgement}

We acknowledge helpful discussions with John Philbin and Xuecheng Tao. This work is primarily supported through the Department of Energy BES QIS program under award number DE-SC0022277 for the work on nonlinearity in low dimensional systems, as well as partially supported by the Quantum Science Center (QSC), a National Quantum Information Science Research Center of the U.S. Department of Energy (DOE) on cavity-control of quantum matter. P.N. is a Moore Inventor Fellow and gratefully acknowledges support through Grant GBMF8048 from the Gordon and Betty Moore Foundation as well as support from a NSF CAREER Award under Grant No. NSF-ECCS-1944085. This research used resources of the National Energy Research Scientific Computing Center, a DOE Office of Science User Facility supported by the Office of Science of the U.S. Department of Energy under Contract No. DE-AC02-05CH11231.

%-----------------------------------------------------------------
%				               Appendix 
%-----------------------------------------------------------------

\appendix

\section{Numerical details {for simulations}}
\label{sup:numerical-details}

In this section, we outline the numerical details {regarding} the coupled light-matter system. First, for the molecular degrees of freedom, we represent {the system} on a uniform real-space grid of $N_{x_{1}} = N_{x_{2}} =150$ grid points with grid spacing $\Delta x_{1} = \Delta x_{2} = 0.1$ atomic units (au) while applying an eighth-order finite-difference scheme for the Laplacian. In the case where we couple the molecular degrees to one photon mode, we sampled 30 photon Fock states in order to correctly represent the coherent state with amplitude $\beta=2$ to obtain converged results. In this case, the dimension of the coupled light-matter system is $150\times150\times 30=675000$.

% Next, in order to simulate the case with coupling explicitly to a photonic bath, we first perform an exact diagonalization of the Hamiltonian of the molecular degrees only and obtain the spectrum of the system (converged eigen-energies $E_{i}$ and eigen-states $|\psi_{i}\rangle$). Now, using the completeness relation $\sum_{i=1}^{\infty}|\psi_{i}\rangle\langle\psi_{i}| = \hat{\mathbb{1}}$, the operators of the molecular system can be expressed as~\cite{loudon2000}
% \begin{align*}
% \hat{H}_{\text{M}} &= \sum_{i=1}E_{i}|\psi_{i}\rangle\langle\psi_{i}|, \;\; 
% \hat{\textbf{R}} = \sum_{i=1}\sum_{j=1} \langle\psi_{i}|\hat{\textbf{R}}|\psi_{j}\rangle |\psi_{i}\rangle\langle\psi_{j}|, \nonumber
% \end{align*}
% where the indices $i,j$ runs over the number of vibrational states considered. For the photonic bath, each photon mode is represented in a basis of Fock number states. To be able to treat the large dimensionality of coupling to a discrete electromagnetic continuum numerically exact, we sample $M = 50$ photon modes where we truncate the Fock space and consider only the vacuum state, the $M$ one-photon states, and the $(M^{2} + M)/2$ two-photon states. Based on the truncation of the Fock number states of the photon modes, the  dimension of the photonic continuum is $1+M+(M^{2}+ M)/2=1326$ and coupling to $20$ lowest vibrational states of each molecule gives a light-matter dimension of $20\times 20\times 30\times 1326=15912000$.

\vspace{10em}


\begin{thebibliography}{47}%
\makeatletter
\providecommand \@ifxundefined [1]{%
 \@ifx{#1\undefined}
}%
\providecommand \@ifnum [1]{%
 \ifnum #1\expandafter \@firstoftwo
 \else \expandafter \@secondoftwo
 \fi
}%
\providecommand \@ifx [1]{%
 \ifx #1\expandafter \@firstoftwo
 \else \expandafter \@secondoftwo
 \fi
}%
\providecommand \natexlab [1]{#1}%
\providecommand \enquote  [1]{``#1''}%
\providecommand \bibnamefont  [1]{#1}%
\providecommand \bibfnamefont [1]{#1}%
\providecommand \citenamefont [1]{#1}%
\providecommand \href@noop [0]{\@secondoftwo}%
\providecommand \href [0]{\begingroup \@sanitize@url \@href}%
\providecommand \@href[1]{\@@startlink{#1}\@@href}%
\providecommand \@@href[1]{\endgroup#1\@@endlink}%
\providecommand \@sanitize@url [0]{\catcode `\\12\catcode `\$12\catcode
  `\&12\catcode `\#12\catcode `\^12\catcode `\_12\catcode `\%12\relax}%
\providecommand \@@startlink[1]{}%
\providecommand \@@endlink[0]{}%
\providecommand \url  [0]{\begingroup\@sanitize@url \@url }%
\providecommand \@url [1]{\endgroup\@href {#1}{\urlprefix }}%
\providecommand \urlprefix  [0]{URL }%
\providecommand \Eprint [0]{\href }%
\providecommand \doibase [0]{https://doi.org/}%
\providecommand \selectlanguage [0]{\@gobble}%
\providecommand \bibinfo  [0]{\@secondoftwo}%
\providecommand \bibfield  [0]{\@secondoftwo}%
\providecommand \translation [1]{[#1]}%
\providecommand \BibitemOpen [0]{}%
\providecommand \bibitemStop [0]{}%
\providecommand \bibitemNoStop [0]{.\EOS\space}%
\providecommand \EOS [0]{\spacefactor3000\relax}%
\providecommand \BibitemShut  [1]{\csname bibitem#1\endcsname}%
\let\auto@bib@innerbib\@empty
%</preamble>
\bibitem [{\citenamefont {Plenio}\ and\ \citenamefont
  {Virmani}(2014)}]{Plenio_Virmani_B_14}%
  \BibitemOpen
  \bibfield  {author} {\bibinfo {author} {\bibfnamefont {M.~B.}\ \bibnamefont
  {Plenio}}\ and\ \bibinfo {author} {\bibfnamefont {S.~S.}\ \bibnamefont
  {Virmani}},\ }\bibfield  {title} {\bibinfo {title} {An introduction to
  entanglement theory},\ }in\ \href
  {https://doi.org/10.1007/978-3-319-04063-9_8} {\emph {\bibinfo {booktitle}
  {Quantum Information and Coherence}}},\ \bibinfo {editor} {edited by\
  \bibinfo {editor} {\bibfnamefont {E.}~\bibnamefont {Andersson}}}\ (\bibinfo
  {publisher} {Springer International Publishing},\ \bibinfo {address} {Cham},\
  \bibinfo {year} {2014})\ pp.\ \bibinfo {pages} {173--209}\BibitemShut
  {NoStop}%
\bibitem [{\citenamefont {McConnell}\ \emph {et~al.}(2015)\citenamefont
  {McConnell}, \citenamefont {Zhang}, \citenamefont {Hu}, \citenamefont
  {\'{C}uk},\ and\ \citenamefont {Vuleti\'{c}}}]{mcConnell2015}%
  \BibitemOpen
  \bibfield  {author} {\bibinfo {author} {\bibfnamefont {R.}~\bibnamefont
  {McConnell}}, \bibinfo {author} {\bibfnamefont {H.}~\bibnamefont {Zhang}},
  \bibinfo {author} {\bibfnamefont {J.}~\bibnamefont {Hu}}, \bibinfo {author}
  {\bibfnamefont {S.}~\bibnamefont {\'{C}uk}},\ and\ \bibinfo {author}
  {\bibfnamefont {V.}~\bibnamefont {Vuleti\'{c}}},\ }\bibfield  {title}
  {\bibinfo {title} {Entanglement with negative wigner function of almost 3,000
  atoms heralded by one photon},\ }\href {https://doi.org/10.1038/nature14293}
  {\bibfield  {journal} {\bibinfo  {journal} {Nature}\ }\textbf {\bibinfo
  {volume} {519}},\ \bibinfo {pages} {439} (\bibinfo {year}
  {2015})}\BibitemShut {NoStop}%
\bibitem [{\citenamefont {Hacker}\ \emph {et~al.}(2019)\citenamefont {Hacker},
  \citenamefont {Welte}, \citenamefont {Daiss}, \citenamefont {Shaukat},
  \citenamefont {Ritter}, \citenamefont {Li},\ and\ \citenamefont
  {Rempe}}]{hacker2019}%
  \BibitemOpen
  \bibfield  {author} {\bibinfo {author} {\bibfnamefont {B.}~\bibnamefont
  {Hacker}}, \bibinfo {author} {\bibfnamefont {S.}~\bibnamefont {Welte}},
  \bibinfo {author} {\bibfnamefont {S.}~\bibnamefont {Daiss}}, \bibinfo
  {author} {\bibfnamefont {A.}~\bibnamefont {Shaukat}}, \bibinfo {author}
  {\bibfnamefont {S.}~\bibnamefont {Ritter}}, \bibinfo {author} {\bibfnamefont
  {L.}~\bibnamefont {Li}},\ and\ \bibinfo {author} {\bibfnamefont
  {G.}~\bibnamefont {Rempe}},\ }\bibfield  {title} {\bibinfo {title}
  {Deterministic creation of entangled atom-light schr\"{o}dinger-cat states},\
  }\href@noop {} {\bibfield  {journal} {\bibinfo  {journal} {Nature Photonics
  13, 110–115}\ } (\bibinfo {year} {2019})}\BibitemShut {NoStop}%
\bibitem [{\citenamefont {Lin}\ \emph {et~al.}(2020)\citenamefont {Lin},
  \citenamefont {Leibrandt}, \citenamefont {Leibfried},\ and\ \citenamefont
  {wen Chou}}]{lin2020}%
  \BibitemOpen
  \bibfield  {author} {\bibinfo {author} {\bibfnamefont {Y.}~\bibnamefont
  {Lin}}, \bibinfo {author} {\bibfnamefont {D.~R.}\ \bibnamefont {Leibrandt}},
  \bibinfo {author} {\bibfnamefont {D.}~\bibnamefont {Leibfried}},\ and\
  \bibinfo {author} {\bibfnamefont {C.}~\bibnamefont {wen Chou}},\ }\bibfield
  {title} {\bibinfo {title} {Quantum entanglement between an atom and a
  molecule},\ }\href {https://doi.org/10.1038/s41586-020-2257-1} {\bibfield
  {journal} {\bibinfo  {journal} {Nature}\ }\textbf {\bibinfo {volume} {581}},\
  \bibinfo {pages} {273} (\bibinfo {year} {2020})}\BibitemShut {NoStop}%
\bibitem [{\citenamefont {DeMille}(2002)}]{deMille2002}%
  \BibitemOpen
  \bibfield  {author} {\bibinfo {author} {\bibfnamefont {D.}~\bibnamefont
  {DeMille}},\ }\bibfield  {title} {\bibinfo {title} {Quantum computation with
  trapped polar molecules},\ }\href
  {https://doi.org/10.1103/PhysRevLett.88.067901} {\bibfield  {journal}
  {\bibinfo  {journal} {Phys. Rev. Lett.}\ }\textbf {\bibinfo {volume} {88}},\
  \bibinfo {pages} {067901} (\bibinfo {year} {2002})}\BibitemShut {NoStop}%
\bibitem [{\citenamefont {Carr}\ \emph {et~al.}(2009)\citenamefont {Carr},
  \citenamefont {DeMille}, \citenamefont {Krems},\ and\ \citenamefont
  {Ye}}]{carr2009}%
  \BibitemOpen
  \bibfield  {author} {\bibinfo {author} {\bibfnamefont {L.~D.}\ \bibnamefont
  {Carr}}, \bibinfo {author} {\bibfnamefont {D.}~\bibnamefont {DeMille}},
  \bibinfo {author} {\bibfnamefont {R.~V.}\ \bibnamefont {Krems}},\ and\
  \bibinfo {author} {\bibfnamefont {J.}~\bibnamefont {Ye}},\ }\bibfield
  {title} {\bibinfo {title} {Cold and ultracold molecules: science, technology
  and applications},\ }\href {https://doi.org/10.1088/1367-2630/11/5/055049}
  {\bibfield  {journal} {\bibinfo  {journal} {New J. Phys.}\ }\textbf {\bibinfo
  {volume} {11}},\ \bibinfo {pages} {055049} (\bibinfo {year}
  {2009})}\BibitemShut {NoStop}%
\bibitem [{\citenamefont {Blackmore}\ \emph {et~al.}(2018)\citenamefont
  {Blackmore}, \citenamefont {Caldwell}, \citenamefont {Gregory}, \citenamefont
  {Bridge}, \citenamefont {Sawant}, \citenamefont {Aldegunde}, \citenamefont
  {Mur-Petit}, \citenamefont {Jaksch}, \citenamefont {Hutson}, \citenamefont
  {Sauer}, \citenamefont {Tarbutt},\ and\ \citenamefont
  {Cornish}}]{blackmore2019}%
  \BibitemOpen
  \bibfield  {author} {\bibinfo {author} {\bibfnamefont {J.~A.}\ \bibnamefont
  {Blackmore}}, \bibinfo {author} {\bibfnamefont {L.}~\bibnamefont {Caldwell}},
  \bibinfo {author} {\bibfnamefont {P.~D.}\ \bibnamefont {Gregory}}, \bibinfo
  {author} {\bibfnamefont {E.~M.}\ \bibnamefont {Bridge}}, \bibinfo {author}
  {\bibfnamefont {R.}~\bibnamefont {Sawant}}, \bibinfo {author} {\bibfnamefont
  {J.}~\bibnamefont {Aldegunde}}, \bibinfo {author} {\bibfnamefont
  {J.}~\bibnamefont {Mur-Petit}}, \bibinfo {author} {\bibfnamefont
  {D.}~\bibnamefont {Jaksch}}, \bibinfo {author} {\bibfnamefont {J.~M.}\
  \bibnamefont {Hutson}}, \bibinfo {author} {\bibfnamefont {B.~E.}\
  \bibnamefont {Sauer}}, \bibinfo {author} {\bibfnamefont {M.~R.}\ \bibnamefont
  {Tarbutt}},\ and\ \bibinfo {author} {\bibfnamefont {S.~L.}\ \bibnamefont
  {Cornish}},\ }\bibfield  {title} {\bibinfo {title} {Ultracold molecules for
  quantum simulation: rotational coherences in caf and rbcs},\ }\href
  {https://doi.org/10.1088/2058-9565/aaee35} {\bibfield  {journal} {\bibinfo
  {journal} {Quantum Science and Technology}\ }\textbf {\bibinfo {volume}
  {4}},\ \bibinfo {pages} {014010} (\bibinfo {year} {2018})}\BibitemShut
  {NoStop}%
\bibitem [{\citenamefont {Ding}\ \emph {et~al.}(2022)\citenamefont {Ding},
  \citenamefont {Knecht}, \citenamefont {Zimborás},\ and\ \citenamefont
  {Schilling}}]{ding2023}%
  \BibitemOpen
  \bibfield  {author} {\bibinfo {author} {\bibfnamefont {L.}~\bibnamefont
  {Ding}}, \bibinfo {author} {\bibfnamefont {S.}~\bibnamefont {Knecht}},
  \bibinfo {author} {\bibfnamefont {Z.}~\bibnamefont {Zimborás}},\ and\
  \bibinfo {author} {\bibfnamefont {C.}~\bibnamefont {Schilling}},\ }\bibfield
  {title} {\bibinfo {title} {Quantum correlations in molecules: from quantum
  resourcing to chemical bonding},\ }\href
  {https://doi.org/10.1088/2058-9565/aca4ee} {\bibfield  {journal} {\bibinfo
  {journal} {Quantum Science and Technology}\ }\textbf {\bibinfo {volume}
  {8}},\ \bibinfo {pages} {015015} (\bibinfo {year} {2022})}\BibitemShut
  {NoStop}%
\bibitem [{\citenamefont {Lenz}\ \emph {et~al.}(2021)\citenamefont {Lenz},
  \citenamefont {K\"{o}nig}, \citenamefont {Hunger},\ and\ \citenamefont {van
  Slageren}}]{lenz2021}%
  \BibitemOpen
  \bibfield  {author} {\bibinfo {author} {\bibfnamefont {S.}~\bibnamefont
  {Lenz}}, \bibinfo {author} {\bibfnamefont {D.}~\bibnamefont {K\"{o}nig}},
  \bibinfo {author} {\bibfnamefont {D.}~\bibnamefont {Hunger}},\ and\ \bibinfo
  {author} {\bibfnamefont {J.}~\bibnamefont {van Slageren}},\ }\bibfield
  {title} {\bibinfo {title} {Room-temperature quantum memories based on
  molecular electron spin ensembles},\ }\href
  {https://doi.org/10.1002/adma.202101673} {\bibfield  {journal} {\bibinfo
  {journal} {Adv.Mater.}\ }\textbf {\bibinfo {volume} {33}},\ \bibinfo {pages}
  {2101673} (\bibinfo {year} {2021})}\BibitemShut {NoStop}%
\bibitem [{\citenamefont {Serrano}\ \emph {et~al.}(2022)\citenamefont
  {Serrano}, \citenamefont {Kuppusamy}, \citenamefont {Heinrich}, \citenamefont
  {Fuhr}, \citenamefont {Hunger}, \citenamefont {Ruben},\ and\ \citenamefont
  {Goldner}}]{serrano2022}%
  \BibitemOpen
  \bibfield  {author} {\bibinfo {author} {\bibfnamefont {D.}~\bibnamefont
  {Serrano}}, \bibinfo {author} {\bibfnamefont {S.~K.}\ \bibnamefont
  {Kuppusamy}}, \bibinfo {author} {\bibfnamefont {B.}~\bibnamefont {Heinrich}},
  \bibinfo {author} {\bibfnamefont {O.}~\bibnamefont {Fuhr}}, \bibinfo {author}
  {\bibfnamefont {D.}~\bibnamefont {Hunger}}, \bibinfo {author} {\bibfnamefont
  {M.}~\bibnamefont {Ruben}},\ and\ \bibinfo {author} {\bibfnamefont
  {P.}~\bibnamefont {Goldner}},\ }\bibfield  {title} {\bibinfo {title}
  {Ultra-narrow optical linewidths in rare-earth molecular crystals},\ }\href
  {https://doi.org/10.1038/s41586-021-04316-2} {\bibfield  {journal} {\bibinfo
  {journal} {Nature}\ }\textbf {\bibinfo {volume} {603}},\ \bibinfo {pages}
  {241–246} (\bibinfo {year} {2022})}\BibitemShut {NoStop}%
\bibitem [{\citenamefont {Albert}\ \emph {et~al.}(2020)\citenamefont {Albert},
  \citenamefont {Covey},\ and\ \citenamefont {Preskill}}]{albert2020}%
  \BibitemOpen
  \bibfield  {author} {\bibinfo {author} {\bibfnamefont {V.~V.}\ \bibnamefont
  {Albert}}, \bibinfo {author} {\bibfnamefont {J.~P.}\ \bibnamefont {Covey}},\
  and\ \bibinfo {author} {\bibfnamefont {J.}~\bibnamefont {Preskill}},\
  }\bibfield  {title} {\bibinfo {title} {Robust encoding of a qubit in a
  molecule},\ }\href {https://doi.org/10.1103/PhysRevX.10.031050} {\bibfield
  {journal} {\bibinfo  {journal} {Phys. Rev. X}\ }\textbf {\bibinfo {volume}
  {10}},\ \bibinfo {pages} {031050} (\bibinfo {year} {2020})}\BibitemShut
  {NoStop}%
\bibitem [{\citenamefont {Hutchison}\ \emph {et~al.}(2012)\citenamefont
  {Hutchison}, \citenamefont {Schwartz}, \citenamefont {Genet}, \citenamefont
  {Devaux},\ and\ \citenamefont {Ebbesen}}]{hutchison2012}%
  \BibitemOpen
  \bibfield  {author} {\bibinfo {author} {\bibfnamefont {J.~A.}\ \bibnamefont
  {Hutchison}}, \bibinfo {author} {\bibfnamefont {T.}~\bibnamefont {Schwartz}},
  \bibinfo {author} {\bibfnamefont {C.}~\bibnamefont {Genet}}, \bibinfo
  {author} {\bibfnamefont {E.}~\bibnamefont {Devaux}},\ and\ \bibinfo {author}
  {\bibfnamefont {T.~W.}\ \bibnamefont {Ebbesen}},\ }\bibfield  {title}
  {\bibinfo {title} {Modifying chemical landscapes by coupling to vacuum
  fields},\ }\href {https://doi.org/10.1002/anie.201107033} {\bibfield
  {journal} {\bibinfo  {journal} {Angewandte Chemie International Edition}\
  }\textbf {\bibinfo {volume} {51}},\ \bibinfo {pages} {1592} (\bibinfo {year}
  {2012})}\BibitemShut {NoStop}%
\bibitem [{\citenamefont {Kowalewski}\ \emph {et~al.}(2016)\citenamefont
  {Kowalewski}, \citenamefont {Bennett},\ and\ \citenamefont
  {Mukamel}}]{kowalewski2016}%
  \BibitemOpen
  \bibfield  {author} {\bibinfo {author} {\bibfnamefont {M.}~\bibnamefont
  {Kowalewski}}, \bibinfo {author} {\bibfnamefont {K.}~\bibnamefont
  {Bennett}},\ and\ \bibinfo {author} {\bibfnamefont {S.}~\bibnamefont
  {Mukamel}},\ }\bibfield  {title} {\bibinfo {title} {Cavity femtochemistry:
  Manipulating nonadiabatic dynamics at avoided crossings},\ }\href
  {https://doi.org/10.1021/acs.jpclett.6b00864} {\bibfield  {journal} {\bibinfo
   {journal} {J. Phys. Chem. Lett.}\ }\textbf {\bibinfo {volume} {7}},\
  \bibinfo {pages} {2050} (\bibinfo {year} {2016})},\ \Eprint
  {https://arxiv.org/abs/http://dx.doi.org/10.1021/acs.jpclett.6b00864}
  {http://dx.doi.org/10.1021/acs.jpclett.6b00864} \BibitemShut {NoStop}%
\bibitem [{\citenamefont {Coles}\ \emph {et~al.}(2014)\citenamefont {Coles},
  \citenamefont {Yang}, \citenamefont {Wang}, \citenamefont {Grant},
  \citenamefont {Taylor}, \citenamefont {Saikin}, \citenamefont {Aspuru-Guzik},
  \citenamefont {Lidzey}, \citenamefont {Tang},\ and\ \citenamefont
  {Smith}}]{coles2014}%
  \BibitemOpen
  \bibfield  {author} {\bibinfo {author} {\bibfnamefont {D.~M.}\ \bibnamefont
  {Coles}}, \bibinfo {author} {\bibfnamefont {Y.}~\bibnamefont {Yang}},
  \bibinfo {author} {\bibfnamefont {Y.}~\bibnamefont {Wang}}, \bibinfo {author}
  {\bibfnamefont {R.~T.}\ \bibnamefont {Grant}}, \bibinfo {author}
  {\bibfnamefont {R.~A.}\ \bibnamefont {Taylor}}, \bibinfo {author}
  {\bibfnamefont {S.~K.}\ \bibnamefont {Saikin}}, \bibinfo {author}
  {\bibfnamefont {A.}~\bibnamefont {Aspuru-Guzik}}, \bibinfo {author}
  {\bibfnamefont {D.~G.}\ \bibnamefont {Lidzey}}, \bibinfo {author}
  {\bibfnamefont {J.~K.-H.}\ \bibnamefont {Tang}},\ and\ \bibinfo {author}
  {\bibfnamefont {J.~M.}\ \bibnamefont {Smith}},\ }\bibfield  {title} {\bibinfo
  {title} {Strong coupling between chlorosomes of photosynthetic bacteria and a
  confined optical cavity mode},\ }\href {https://doi.org/10.1038/ncomms6561}
  {\bibfield  {journal} {\bibinfo  {journal} {Nature Communications}\ }\textbf
  {\bibinfo {volume} {5}},\ \bibinfo {pages} {5561} (\bibinfo {year}
  {2014})}\BibitemShut {NoStop}%
\bibitem [{\citenamefont {Zhong}\ \emph {et~al.}(2016)\citenamefont {Zhong},
  \citenamefont {Chervy}, \citenamefont {Wang}, \citenamefont {George},
  \citenamefont {Thomas}, \citenamefont {Hutchison}, \citenamefont {Devaux},
  \citenamefont {Genet},\ and\ \citenamefont {Ebbesen}}]{zhong2016}%
  \BibitemOpen
  \bibfield  {author} {\bibinfo {author} {\bibfnamefont {X.}~\bibnamefont
  {Zhong}}, \bibinfo {author} {\bibfnamefont {T.}~\bibnamefont {Chervy}},
  \bibinfo {author} {\bibfnamefont {S.}~\bibnamefont {Wang}}, \bibinfo {author}
  {\bibfnamefont {J.}~\bibnamefont {George}}, \bibinfo {author} {\bibfnamefont
  {A.}~\bibnamefont {Thomas}}, \bibinfo {author} {\bibfnamefont {J.~A.}\
  \bibnamefont {Hutchison}}, \bibinfo {author} {\bibfnamefont {E.}~\bibnamefont
  {Devaux}}, \bibinfo {author} {\bibfnamefont {C.}~\bibnamefont {Genet}},\ and\
  \bibinfo {author} {\bibfnamefont {T.~W.}\ \bibnamefont {Ebbesen}},\
  }\bibfield  {title} {\bibinfo {title} {Non-radiative energy transfer mediated
  by hybrid light-matter states},\ }\href
  {https://doi.org/10.1002/anie.201600428} {\bibfield  {journal} {\bibinfo
  {journal} {Angewandte Chemie International Edition}\ }\textbf {\bibinfo
  {volume} {55}},\ \bibinfo {pages} {6202} (\bibinfo {year}
  {2016})}\BibitemShut {NoStop}%
\bibitem [{\citenamefont {Chervy}\ \emph {et~al.}(2016)\citenamefont {Chervy},
  \citenamefont {Xu}, \citenamefont {Duan}, \citenamefont {Wang}, \citenamefont
  {Mager}, \citenamefont {Frerejean}, \citenamefont {M\"{u}nninghoff},
  \citenamefont {Tinnemans}, \citenamefont {Hutchison}, \citenamefont {Genet},
  \citenamefont {Rowan}, \citenamefont {Rasing},\ and\ \citenamefont
  {Ebbesen}}]{chervy2016}%
  \BibitemOpen
  \bibfield  {author} {\bibinfo {author} {\bibfnamefont {T.}~\bibnamefont
  {Chervy}}, \bibinfo {author} {\bibfnamefont {J.}~\bibnamefont {Xu}}, \bibinfo
  {author} {\bibfnamefont {Y.}~\bibnamefont {Duan}}, \bibinfo {author}
  {\bibfnamefont {C.}~\bibnamefont {Wang}}, \bibinfo {author} {\bibfnamefont
  {L.}~\bibnamefont {Mager}}, \bibinfo {author} {\bibfnamefont
  {M.}~\bibnamefont {Frerejean}}, \bibinfo {author} {\bibfnamefont {J.~A.~W.}\
  \bibnamefont {M\"{u}nninghoff}}, \bibinfo {author} {\bibfnamefont
  {P.}~\bibnamefont {Tinnemans}}, \bibinfo {author} {\bibfnamefont {J.~A.}\
  \bibnamefont {Hutchison}}, \bibinfo {author} {\bibfnamefont {C.}~\bibnamefont
  {Genet}}, \bibinfo {author} {\bibfnamefont {A.~E.}\ \bibnamefont {Rowan}},
  \bibinfo {author} {\bibfnamefont {T.}~\bibnamefont {Rasing}},\ and\ \bibinfo
  {author} {\bibfnamefont {T.~W.}\ \bibnamefont {Ebbesen}},\ }\bibfield
  {title} {\bibinfo {title} {High-eﬃciency second-harmonic generation from
  hybrid light-matter states},\ }\href
  {https://doi.org/10.1021/acs.nanolett.6b02567} {\bibfield  {journal}
  {\bibinfo  {journal} {Nano Lett.}\ }\textbf {\bibinfo {volume} {16}},\
  \bibinfo {pages} {7352} (\bibinfo {year} {2016})}\BibitemShut {NoStop}%
\bibitem [{\citenamefont {Barachati}\ \emph {et~al.}(2018)\citenamefont
  {Barachati}, \citenamefont {Simon}, \citenamefont {Getmanenko}, \citenamefont
  {Barlow}, \citenamefont {Marder},\ and\ \citenamefont
  {K\'{e}na-Cohen}}]{barachati2018}%
  \BibitemOpen
  \bibfield  {author} {\bibinfo {author} {\bibfnamefont {F.}~\bibnamefont
  {Barachati}}, \bibinfo {author} {\bibfnamefont {J.}~\bibnamefont {Simon}},
  \bibinfo {author} {\bibfnamefont {Y.~A.}\ \bibnamefont {Getmanenko}},
  \bibinfo {author} {\bibfnamefont {S.}~\bibnamefont {Barlow}}, \bibinfo
  {author} {\bibfnamefont {S.~R.}\ \bibnamefont {Marder}},\ and\ \bibinfo
  {author} {\bibfnamefont {S.}~\bibnamefont {K\'{e}na-Cohen}},\ }\bibfield
  {title} {\bibinfo {title} {Tunable third-harmonic generation from polaritons
  in theultrastrong coupling regime},\ }\href
  {https://doi.org/10.1021/acsphotonics.7b00305} {\bibfield  {journal}
  {\bibinfo  {journal} {ACS Photonics}\ }\textbf {\bibinfo {volume} {5}},\
  \bibinfo {pages} {119} (\bibinfo {year} {2018})}\BibitemShut {NoStop}%
\bibitem [{\citenamefont {Welakuh}\ and\ \citenamefont
  {Narang}(2023)}]{welakuh2022b}%
  \BibitemOpen
  \bibfield  {author} {\bibinfo {author} {\bibfnamefont {D.~M.}\ \bibnamefont
  {Welakuh}}\ and\ \bibinfo {author} {\bibfnamefont {P.}~\bibnamefont
  {Narang}},\ }\bibfield  {title} {\bibinfo {title} {Tunable and efficient
  harmonic generation from strongly coupled light-matter system},\ }\href
  {https://doi.org/10.1021/acsphotonics.2c00966} {\bibfield  {journal}
  {\bibinfo  {journal} {ACS Photonics}\ }\textbf {\bibinfo {volume} {10}},\
  \bibinfo {pages} {383–393} (\bibinfo {year} {2023})}\BibitemShut {NoStop}%
\bibitem [{\citenamefont {Welakuh}\ and\ \citenamefont
  {Narang}(2022)}]{welakuh2022c}%
  \BibitemOpen
  \bibfield  {author} {\bibinfo {author} {\bibfnamefont {D.~M.}\ \bibnamefont
  {Welakuh}}\ and\ \bibinfo {author} {\bibfnamefont {P.}~\bibnamefont
  {Narang}},\ }\bibfield  {title} {\bibinfo {title} {Nonlinear optical
  processes in centrosymmetric systems by strong-coupling-induced symmetry
  breaking},\ }\href@noop {} {\bibfield  {journal} {\bibinfo  {journal} {arxiv
  (to be submitted)}\ } (\bibinfo {year} {2022})}\BibitemShut {NoStop}%
\bibitem [{\citenamefont {Thompson}\ \emph {et~al.}(2006)\citenamefont
  {Thompson}, \citenamefont {Simon}, \citenamefont {Loh},\ and\ \citenamefont
  {Vuleti\'{c}}}]{thompson2006}%
  \BibitemOpen
  \bibfield  {author} {\bibinfo {author} {\bibfnamefont {J.~K.}\ \bibnamefont
  {Thompson}}, \bibinfo {author} {\bibfnamefont {J.}~\bibnamefont {Simon}},
  \bibinfo {author} {\bibfnamefont {H.}~\bibnamefont {Loh}},\ and\ \bibinfo
  {author} {\bibfnamefont {V.}~\bibnamefont {Vuleti\'{c}}},\ }\bibfield
  {title} {\bibinfo {title} {A high-brightness source of narrowband,
  identical-photon pairs},\ }\href {https://doi.org/10.1126/science.1127676}
  {\bibfield  {journal} {\bibinfo  {journal} {Science}\ }\textbf {\bibinfo
  {volume} {313}},\ \bibinfo {pages} {74} (\bibinfo {year} {2006})}\BibitemShut
  {NoStop}%
\bibitem [{\citenamefont {Thomas}\ \emph {et~al.}(2019)\citenamefont {Thomas},
  \citenamefont {Lethuillier-Karl}, \citenamefont {Nagarajan}, \citenamefont
  {Vergauwe}, \citenamefont {George}, \citenamefont {Chervy}, \citenamefont
  {Shalabney}, \citenamefont {Devaux}, \citenamefont {Genet}, \citenamefont
  {Moran},\ and\ \citenamefont {Ebbesen}}]{thomas2019}%
  \BibitemOpen
  \bibfield  {author} {\bibinfo {author} {\bibfnamefont {A.}~\bibnamefont
  {Thomas}}, \bibinfo {author} {\bibfnamefont {L.}~\bibnamefont
  {Lethuillier-Karl}}, \bibinfo {author} {\bibfnamefont {K.}~\bibnamefont
  {Nagarajan}}, \bibinfo {author} {\bibfnamefont {R.~M.~A.}\ \bibnamefont
  {Vergauwe}}, \bibinfo {author} {\bibfnamefont {J.}~\bibnamefont {George}},
  \bibinfo {author} {\bibfnamefont {T.}~\bibnamefont {Chervy}}, \bibinfo
  {author} {\bibfnamefont {A.}~\bibnamefont {Shalabney}}, \bibinfo {author}
  {\bibfnamefont {E.}~\bibnamefont {Devaux}}, \bibinfo {author} {\bibfnamefont
  {C.}~\bibnamefont {Genet}}, \bibinfo {author} {\bibfnamefont
  {J.}~\bibnamefont {Moran}},\ and\ \bibinfo {author} {\bibfnamefont {T.~W.}\
  \bibnamefont {Ebbesen}},\ }\bibfield  {title} {\bibinfo {title} {Tilting a
  ground-state reactivity landscape by vibrational strong coupling},\ }\href
  {https://doi.org/10.1126/science.aau7742} {\bibfield  {journal} {\bibinfo
  {journal} {Science}\ }\textbf {\bibinfo {volume} {363}},\ \bibinfo {pages}
  {615} (\bibinfo {year} {2019})}\BibitemShut {NoStop}%
\bibitem [{\citenamefont {Latini}\ \emph {et~al.}(2021)\citenamefont {Latini},
  \citenamefont {Shin}, \citenamefont {Sato}, \citenamefont {Sch\"{a}fer},
  \citenamefont {Giovannini}, \citenamefont {H\"{u}bener},\ and\ \citenamefont
  {Rubio}}]{latini2021}%
  \BibitemOpen
  \bibfield  {author} {\bibinfo {author} {\bibfnamefont {S.}~\bibnamefont
  {Latini}}, \bibinfo {author} {\bibfnamefont {D.}~\bibnamefont {Shin}},
  \bibinfo {author} {\bibfnamefont {S.~A.}\ \bibnamefont {Sato}}, \bibinfo
  {author} {\bibfnamefont {C.}~\bibnamefont {Sch\"{a}fer}}, \bibinfo {author}
  {\bibfnamefont {U.~D.}\ \bibnamefont {Giovannini}}, \bibinfo {author}
  {\bibfnamefont {H.}~\bibnamefont {H\"{u}bener}},\ and\ \bibinfo {author}
  {\bibfnamefont {A.}~\bibnamefont {Rubio}},\ }\bibfield  {title} {\bibinfo
  {title} {The ferroelectric photo ground state of srtio<sub>3</sub>: Cavity
  materials engineering},\ }\href {https://doi.org/10.1073/pnas.2105618118}
  {\bibfield  {journal} {\bibinfo  {journal} {Proceedings of the National
  Academy of Sciences}\ }\textbf {\bibinfo {volume} {118}},\ \bibinfo {pages}
  {e2105618118} (\bibinfo {year} {2021})}\BibitemShut {NoStop}%
\bibitem [{\citenamefont {Liu}\ \emph {et~al.}(2015)\citenamefont {Liu},
  \citenamefont {Galfsky}, \citenamefont {Sun}, \citenamefont {Xia},
  \citenamefont {chen Lin}, \citenamefont {Lee}, \citenamefont
  {K\'{e}na-Cohen},\ and\ \citenamefont {Menon}}]{liu2015}%
  \BibitemOpen
  \bibfield  {author} {\bibinfo {author} {\bibfnamefont {X.}~\bibnamefont
  {Liu}}, \bibinfo {author} {\bibfnamefont {T.}~\bibnamefont {Galfsky}},
  \bibinfo {author} {\bibfnamefont {Z.}~\bibnamefont {Sun}}, \bibinfo {author}
  {\bibfnamefont {F.}~\bibnamefont {Xia}}, \bibinfo {author} {\bibfnamefont
  {E.}~\bibnamefont {chen Lin}}, \bibinfo {author} {\bibfnamefont {Y.-H.}\
  \bibnamefont {Lee}}, \bibinfo {author} {\bibfnamefont {S.}~\bibnamefont
  {K\'{e}na-Cohen}},\ and\ \bibinfo {author} {\bibfnamefont {V.~M.}\
  \bibnamefont {Menon}},\ }\bibfield  {title} {\bibinfo {title} {Strong
  light–matter coupling in two-dimensional atomic crystals},\ }\href
  {https://doi.org/10.1038/nphoton.2014.304} {\bibfield  {journal} {\bibinfo
  {journal} {Nature Photonics}\ }\textbf {\bibinfo {volume} {9}},\ \bibinfo
  {pages} {1749} (\bibinfo {year} {2015})}\BibitemShut {NoStop}%
\bibitem [{\citenamefont {Thomas}\ \emph {et~al.}(2016)\citenamefont {Thomas},
  \citenamefont {George}, \citenamefont {Shalabney}, \citenamefont {Dryzhakov},
  \citenamefont {Varma}, \citenamefont {Moran}, \citenamefont {Chervy},
  \citenamefont {Zhong}, \citenamefont {Devaux}, \citenamefont {Genet},
  \citenamefont {Hutchison},\ and\ \citenamefont {Ebbesen}}]{thomas2016}%
  \BibitemOpen
  \bibfield  {author} {\bibinfo {author} {\bibfnamefont {A.}~\bibnamefont
  {Thomas}}, \bibinfo {author} {\bibfnamefont {J.}~\bibnamefont {George}},
  \bibinfo {author} {\bibfnamefont {A.}~\bibnamefont {Shalabney}}, \bibinfo
  {author} {\bibfnamefont {M.}~\bibnamefont {Dryzhakov}}, \bibinfo {author}
  {\bibfnamefont {S.~J.}\ \bibnamefont {Varma}}, \bibinfo {author}
  {\bibfnamefont {J.}~\bibnamefont {Moran}}, \bibinfo {author} {\bibfnamefont
  {T.}~\bibnamefont {Chervy}}, \bibinfo {author} {\bibfnamefont
  {X.}~\bibnamefont {Zhong}}, \bibinfo {author} {\bibfnamefont
  {E.}~\bibnamefont {Devaux}}, \bibinfo {author} {\bibfnamefont
  {C.}~\bibnamefont {Genet}}, \bibinfo {author} {\bibfnamefont {J.~A.}\
  \bibnamefont {Hutchison}},\ and\ \bibinfo {author} {\bibfnamefont {T.~W.}\
  \bibnamefont {Ebbesen}},\ }\bibfield  {title} {\bibinfo {title} {Ground-state
  chemical reactivity under vibrational coupling to the vacuum electromagnetic
  field},\ }\href {https://doi.org/10.1002/anie.201605504} {\bibfield
  {journal} {\bibinfo  {journal} {Angewandte Chemie International Edition}\
  }\textbf {\bibinfo {volume} {55}},\ \bibinfo {pages} {11462} (\bibinfo {year}
  {2016})}\BibitemShut {NoStop}%
\bibitem [{\citenamefont {Lather}\ and\ \citenamefont
  {George}(2020)}]{lather2020}%
  \BibitemOpen
  \bibfield  {author} {\bibinfo {author} {\bibfnamefont {J.}~\bibnamefont
  {Lather}}\ and\ \bibinfo {author} {\bibfnamefont {J.}~\bibnamefont
  {George}},\ }\bibfield  {title} {\bibinfo {title} {Improving enzyme catalytic
  efficiency by co-operative vibrational strong coupling of water},\ }\href
  {https://doi.org/10.1021/acs.jpclett.0c03003} {\bibfield  {journal} {\bibinfo
   {journal} {J. Phys. Chem. Lett.}\ }\textbf {\bibinfo {volume} {12}},\
  \bibinfo {pages} {379} (\bibinfo {year} {2020})}\BibitemShut {NoStop}%
\bibitem [{\citenamefont {Sidler}\ \emph {et~al.}(2023)\citenamefont {Sidler},
  \citenamefont {Ruggenthaler},\ and\ \citenamefont {Rubio}}]{sidler2023}%
  \BibitemOpen
  \bibfield  {author} {\bibinfo {author} {\bibfnamefont {D.}~\bibnamefont
  {Sidler}}, \bibinfo {author} {\bibfnamefont {M.}~\bibnamefont
  {Ruggenthaler}},\ and\ \bibinfo {author} {\bibfnamefont {A.}~\bibnamefont
  {Rubio}},\ }\bibfield  {title} {\bibinfo {title} {Exact solution for a real
  polaritonic system under vibrational strong coupling in thermodynamic
  equilibrium: Absence of zero temperature and loss of light-matter
  entanglement},\ }\href {https://doi.org/10.48550/arXiv.2208.01326} {\bibfield
   {journal} {\bibinfo  {journal} {arXiv:2208.01326}\ } (\bibinfo {year}
  {2023})}\BibitemShut {NoStop}%
\bibitem [{\citenamefont {Wellnitz}\ \emph {et~al.}(2022)\citenamefont
  {Wellnitz}, \citenamefont {Pupillo},\ and\ \citenamefont
  {Schachenmayer}}]{wellnitz2022}%
  \BibitemOpen
  \bibfield  {author} {\bibinfo {author} {\bibfnamefont {D.}~\bibnamefont
  {Wellnitz}}, \bibinfo {author} {\bibfnamefont {G.}~\bibnamefont {Pupillo}},\
  and\ \bibinfo {author} {\bibfnamefont {J.}~\bibnamefont {Schachenmayer}},\
  }\bibfield  {title} {\bibinfo {title} {Disorder enhanced vibrational
  entanglement and dynamics in polaritonic chemistry},\ }\href
  {https://doi.org/10.1038/s42005-022-00892-5} {\bibfield  {journal} {\bibinfo
  {journal} {Communications Physics}\ }\textbf {\bibinfo {volume} {5}},\
  \bibinfo {pages} {120} (\bibinfo {year} {2022})}\BibitemShut {NoStop}%
\bibitem [{\citenamefont {Fischer}\ and\ \citenamefont
  {Saalfrank}(2023)}]{fischer2023}%
  \BibitemOpen
  \bibfield  {author} {\bibinfo {author} {\bibfnamefont {E.~W.}\ \bibnamefont
  {Fischer}}\ and\ \bibinfo {author} {\bibfnamefont {P.}~\bibnamefont
  {Saalfrank}},\ }\bibfield  {title} {\bibinfo {title} {Cavity-catalyzed
  hydrogen transfer dynamics in an entangled molecular ensemble under
  vibrational strong coupling},\ }\href {https://doi.org/10.1039/D3CP00175J}
  {\bibfield  {journal} {\bibinfo  {journal} {Phys. Chem. Chem. Phys.}\ }
  (\bibinfo {year} {2023})}\BibitemShut {NoStop}%
\bibitem [{\citenamefont {Cohen‐Tannoudji}\ \emph {et~al.}(1989)\citenamefont
  {Cohen‐Tannoudji}, \citenamefont {Dupont‐Roc},\ and\ \citenamefont
  {Grynberg}}]{tannoudji1989}%
  \BibitemOpen
  \bibfield  {author} {\bibinfo {author} {\bibfnamefont {C.}~\bibnamefont
  {Cohen‐Tannoudji}}, \bibinfo {author} {\bibfnamefont {J.}~\bibnamefont
  {Dupont‐Roc}},\ and\ \bibinfo {author} {\bibfnamefont {G.}~\bibnamefont
  {Grynberg}},\ }\href
  {https://www.wiley.com/en-us/Photons+and+Atoms%3A+Introduction+to+Quantum+Electrodynamics-p-9780471184331}
  {\emph {\bibinfo {title} {Photons and Atoms: Introduction to Quantum
  Electrodynamics}}}\ (\bibinfo  {publisher} {John Wiley \& Sons, Inc.},\
  \bibinfo {year} {1989})\BibitemShut {NoStop}%
\bibitem [{\citenamefont {Spohn}(2004)}]{spohn2004}%
  \BibitemOpen
  \bibfield  {author} {\bibinfo {author} {\bibfnamefont {H.}~\bibnamefont
  {Spohn}},\ }\href
  {https://www.cambridge.org/core/books/dynamics-of-charged-particles-and-their-radiation-field/92D241C60F65E559EB1B7AFCD9E47F43}
  {\emph {\bibinfo {title} {Dynamics of charged particles and their radiation
  field}}}\ (\bibinfo  {publisher} {Cambridge university press},\ \bibinfo
  {year} {2004})\BibitemShut {NoStop}%
\bibitem [{\citenamefont {Rokaj}\ \emph {et~al.}(2018)\citenamefont {Rokaj},
  \citenamefont {Welakuh}, \citenamefont {Ruggenthaler},\ and\ \citenamefont
  {Rubio}}]{rokaj2017}%
  \BibitemOpen
  \bibfield  {author} {\bibinfo {author} {\bibfnamefont {V.}~\bibnamefont
  {Rokaj}}, \bibinfo {author} {\bibfnamefont {D.~M.}\ \bibnamefont {Welakuh}},
  \bibinfo {author} {\bibfnamefont {M.}~\bibnamefont {Ruggenthaler}},\ and\
  \bibinfo {author} {\bibfnamefont {A.}~\bibnamefont {Rubio}},\ }\bibfield
  {title} {\bibinfo {title} {Light–matter interaction in the long-wavelength
  limit: no ground-state without dipole self-energy},\ }\href
  {http://stacks.iop.org/0953-4075/51/i=3/a=034005} {\bibfield  {journal}
  {\bibinfo  {journal} {Journal of Physics B: Atomic, Molecular and Optical
  Physics}\ }\textbf {\bibinfo {volume} {51}},\ \bibinfo {pages} {034005}
  (\bibinfo {year} {2018})}\BibitemShut {NoStop}%
\bibitem [{\citenamefont {Flick}\ \emph {et~al.}(2019)\citenamefont {Flick},
  \citenamefont {Welakuh}, \citenamefont {Ruggenthaler}, \citenamefont
  {Appel},\ and\ \citenamefont {Rubio}}]{flick2019}%
  \BibitemOpen
  \bibfield  {author} {\bibinfo {author} {\bibfnamefont {J.}~\bibnamefont
  {Flick}}, \bibinfo {author} {\bibfnamefont {D.~M.}\ \bibnamefont {Welakuh}},
  \bibinfo {author} {\bibfnamefont {M.}~\bibnamefont {Ruggenthaler}}, \bibinfo
  {author} {\bibfnamefont {H.}~\bibnamefont {Appel}},\ and\ \bibinfo {author}
  {\bibfnamefont {A.}~\bibnamefont {Rubio}},\ }\bibfield  {title} {\bibinfo
  {title} {Light-matter response in nonrelativistic quantum electrodynamics},\
  }\href {https://doi.org/10.1021/acsphotonics.9b00768} {\bibfield  {journal}
  {\bibinfo  {journal} {ACS Photonics}\ }\textbf {\bibinfo {volume} {6}},\
  \bibinfo {pages} {2757} (\bibinfo {year} {2019})}\BibitemShut {NoStop}%
\bibitem [{\citenamefont {Kockum}\ \emph {et~al.}(2019)\citenamefont {Kockum},
  \citenamefont {Miranowicz}, \citenamefont {Liberato}, \citenamefont
  {Savasta},\ and\ \citenamefont {Nori}}]{kockum2019}%
  \BibitemOpen
  \bibfield  {author} {\bibinfo {author} {\bibfnamefont {A.~F.}\ \bibnamefont
  {Kockum}}, \bibinfo {author} {\bibfnamefont {A.}~\bibnamefont {Miranowicz}},
  \bibinfo {author} {\bibfnamefont {S.~D.}\ \bibnamefont {Liberato}}, \bibinfo
  {author} {\bibfnamefont {S.}~\bibnamefont {Savasta}},\ and\ \bibinfo {author}
  {\bibfnamefont {F.}~\bibnamefont {Nori}},\ }\bibfield  {title} {\bibinfo
  {title} {Ultrastrong coupling between light and matter},\ }\href
  {https://doi.org/10.1038/s42254-018-0006-2} {\bibfield  {journal} {\bibinfo
  {journal} {Nature Reviews Physics}\ }\textbf {\bibinfo {volume} {1}},\
  \bibinfo {pages} {19} (\bibinfo {year} {2019})}\BibitemShut {NoStop}%
\bibitem [{\citenamefont {Wernsdorfer}\ and\ \citenamefont
  {Ruben}(2019)}]{wernsdorfer2019}%
  \BibitemOpen
  \bibfield  {author} {\bibinfo {author} {\bibfnamefont {W.}~\bibnamefont
  {Wernsdorfer}}\ and\ \bibinfo {author} {\bibfnamefont {M.}~\bibnamefont
  {Ruben}},\ }\bibfield  {title} {\bibinfo {title} {Synthetic hilbert space
  engineering of molecular qudits: Isotopologue chemistry},\ }\href
  {https://doi.org/10.1002/adma.201806687} {\bibfield  {journal} {\bibinfo
  {journal} {Adv.Mater.}\ }\textbf {\bibinfo {volume} {31}},\ \bibinfo {pages}
  {1806687} (\bibinfo {year} {2019})}\BibitemShut {NoStop}%
\bibitem [{\citenamefont {Wasielewski}\ \emph {et~al.}(2020)\citenamefont
  {Wasielewski}, \citenamefont {Forbes}, \citenamefont {Frank}, \citenamefont
  {Kowalski}, \citenamefont {Scholes}, \citenamefont {Yuen-Zhou}, \citenamefont
  {Baldo}, \citenamefont {Freedman}, \citenamefont {Goldsmith}, \citenamefont
  {III}, \citenamefont {Kirk}, \citenamefont {McCusker}, \citenamefont
  {Ogilvie}, \citenamefont {Shultz}, \citenamefont {Stoll},\ and\ \citenamefont
  {Whaley}}]{wasielewski2020}%
  \BibitemOpen
  \bibfield  {author} {\bibinfo {author} {\bibfnamefont {M.~R.}\ \bibnamefont
  {Wasielewski}}, \bibinfo {author} {\bibfnamefont {M.~D.~E.}\ \bibnamefont
  {Forbes}}, \bibinfo {author} {\bibfnamefont {N.~L.}\ \bibnamefont {Frank}},
  \bibinfo {author} {\bibfnamefont {K.}~\bibnamefont {Kowalski}}, \bibinfo
  {author} {\bibfnamefont {G.~D.}\ \bibnamefont {Scholes}}, \bibinfo {author}
  {\bibfnamefont {J.}~\bibnamefont {Yuen-Zhou}}, \bibinfo {author}
  {\bibfnamefont {M.~A.}\ \bibnamefont {Baldo}}, \bibinfo {author}
  {\bibfnamefont {D.~E.}\ \bibnamefont {Freedman}}, \bibinfo {author}
  {\bibfnamefont {R.~H.}\ \bibnamefont {Goldsmith}}, \bibinfo {author}
  {\bibfnamefont {T.~G.}\ \bibnamefont {III}}, \bibinfo {author} {\bibfnamefont
  {M.~L.}\ \bibnamefont {Kirk}}, \bibinfo {author} {\bibfnamefont {J.~K.}\
  \bibnamefont {McCusker}}, \bibinfo {author} {\bibfnamefont {J.~P.}\
  \bibnamefont {Ogilvie}}, \bibinfo {author} {\bibfnamefont {D.~A.}\
  \bibnamefont {Shultz}}, \bibinfo {author} {\bibfnamefont {S.}~\bibnamefont
  {Stoll}},\ and\ \bibinfo {author} {\bibfnamefont {K.~B.}\ \bibnamefont
  {Whaley}},\ }\bibfield  {title} {\bibinfo {title} {Exploiting chemistry and
  molecular systems for quantum information science},\ }\href
  {https://doi.org/10.1038/s41570-020-0200-5} {\bibfield  {journal} {\bibinfo
  {journal} {Nature Reviews Chemistry}\ }\textbf {\bibinfo {volume} {4}},\
  \bibinfo {pages} {490–504} (\bibinfo {year} {2020})}\BibitemShut {NoStop}%
\bibitem [{\citenamefont {Korolkov}\ \emph {et~al.}(1996)\citenamefont
  {Korolkov}, \citenamefont {Paramonov},\ and\ \citenamefont
  {Schmidt}}]{korolkov1996}%
  \BibitemOpen
  \bibfield  {author} {\bibinfo {author} {\bibfnamefont {M.~V.}\ \bibnamefont
  {Korolkov}}, \bibinfo {author} {\bibfnamefont {G.~K.}\ \bibnamefont
  {Paramonov}},\ and\ \bibinfo {author} {\bibfnamefont {B.}~\bibnamefont
  {Schmidt}},\ }\bibfield  {title} {\bibinfo {title} {State‐selective control
  for vibrational excitation and dissociation of diatomic molecules with shaped
  ultrashort infrared laser pulses},\ }\href {https://doi.org/10.1063/1.472058}
  {\bibfield  {journal} {\bibinfo  {journal} {J. Chem. Phys.}\ }\textbf
  {\bibinfo {volume} {105}},\ \bibinfo {pages} {1862} (\bibinfo {year}
  {1996})}\BibitemShut {NoStop}%
\bibitem [{\citenamefont {Paramonov}\ and\ \citenamefont
  {Saalfrank}(2009)}]{paramonov2009}%
  \BibitemOpen
  \bibfield  {author} {\bibinfo {author} {\bibfnamefont {G.~K.}\ \bibnamefont
  {Paramonov}}\ and\ \bibinfo {author} {\bibfnamefont {P.}~\bibnamefont
  {Saalfrank}},\ }\bibfield  {title} {\bibinfo {title} {Time-evolution operator
  method for non-markovian density matrix propagation in time and space
  representation: Application to laser association of oh in an environment},\
  }\href {https://doi.org/10.1103/PhysRevA.79.013415} {\bibfield  {journal}
  {\bibinfo  {journal} {Phys. Rev. A}\ }\textbf {\bibinfo {volume} {79}},\
  \bibinfo {pages} {013415} (\bibinfo {year} {2009})}\BibitemShut {NoStop}%
\bibitem [{\citenamefont {Mecke}(1950)}]{mecke1950}%
  \BibitemOpen
  \bibfield  {author} {\bibinfo {author} {\bibfnamefont {R.}~\bibnamefont
  {Mecke}},\ }\href@noop {} {\bibfield  {journal} {\bibinfo  {journal}
  {Zeitschrift f\"{u}r Elektrochemie}\ }\textbf {\bibinfo {volume} {54}},\
  \bibinfo {pages} {38} (\bibinfo {year} {1950})}\BibitemShut {NoStop}%
\bibitem [{\citenamefont {Hammerer}\ \emph {et~al.}(2010)\citenamefont
  {Hammerer}, \citenamefont {S\o{}rensen},\ and\ \citenamefont
  {Polzik}}]{hammerer2010}%
  \BibitemOpen
  \bibfield  {author} {\bibinfo {author} {\bibfnamefont {K.}~\bibnamefont
  {Hammerer}}, \bibinfo {author} {\bibfnamefont {A.~S.}\ \bibnamefont
  {S\o{}rensen}},\ and\ \bibinfo {author} {\bibfnamefont {E.~S.}\ \bibnamefont
  {Polzik}},\ }\bibfield  {title} {\bibinfo {title} {Quantum interface between
  light and atomic ensembles},\ }\href
  {https://doi.org/10.1103/RevModPhys.82.1041} {\bibfield  {journal} {\bibinfo
  {journal} {Rev. Mod. Phys.}\ }\textbf {\bibinfo {volume} {82}},\ \bibinfo
  {pages} {1041} (\bibinfo {year} {2010})}\BibitemShut {NoStop}%
\bibitem [{\citenamefont {Bennett}\ \emph {et~al.}(1996)\citenamefont
  {Bennett}, \citenamefont {DiVincenzo}, \citenamefont {Smolin},\ and\
  \citenamefont {Wootters}}]{Bennett_etal_PRA_96}%
  \BibitemOpen
  \bibfield  {author} {\bibinfo {author} {\bibfnamefont {C.~H.}\ \bibnamefont
  {Bennett}}, \bibinfo {author} {\bibfnamefont {D.~P.}\ \bibnamefont
  {DiVincenzo}}, \bibinfo {author} {\bibfnamefont {J.~A.}\ \bibnamefont
  {Smolin}},\ and\ \bibinfo {author} {\bibfnamefont {W.~K.}\ \bibnamefont
  {Wootters}},\ }\bibfield  {title} {\bibinfo {title} {Mixed-state entanglement
  and quantum error correction},\ }\href
  {https://doi.org/10.1103/PhysRevA.54.3824} {\bibfield  {journal} {\bibinfo
  {journal} {Phys. Rev. A}\ }\textbf {\bibinfo {volume} {54}},\ \bibinfo
  {pages} {3824} (\bibinfo {year} {1996})}\BibitemShut {NoStop}%
\bibitem [{\citenamefont {Plenio}(2005)}]{plenio2005}%
  \BibitemOpen
  \bibfield  {author} {\bibinfo {author} {\bibfnamefont {M.~B.}\ \bibnamefont
  {Plenio}},\ }\bibfield  {title} {\bibinfo {title} {Logarithmic negativity: A
  full entanglement monotone that is not convex},\ }\href
  {https://doi.org/10.1103/PhysRevLett.95.090503} {\bibfield  {journal}
  {\bibinfo  {journal} {Phys. Rev. Lett.}\ }\textbf {\bibinfo {volume} {95}},\
  \bibinfo {pages} {090503} (\bibinfo {year} {2005})}\BibitemShut {NoStop}%
\bibitem [{\citenamefont {Zhong}\ \emph {et~al.}(2017)\citenamefont {Zhong},
  \citenamefont {Chervy}, \citenamefont {Zhang}, \citenamefont {Thomas},
  \citenamefont {George}, \citenamefont {Genet}, \citenamefont {Hutchison},\
  and\ \citenamefont {Ebbesen}}]{zhong2017}%
  \BibitemOpen
  \bibfield  {author} {\bibinfo {author} {\bibfnamefont {X.}~\bibnamefont
  {Zhong}}, \bibinfo {author} {\bibfnamefont {T.}~\bibnamefont {Chervy}},
  \bibinfo {author} {\bibfnamefont {L.}~\bibnamefont {Zhang}}, \bibinfo
  {author} {\bibfnamefont {A.}~\bibnamefont {Thomas}}, \bibinfo {author}
  {\bibfnamefont {J.}~\bibnamefont {George}}, \bibinfo {author} {\bibfnamefont
  {C.}~\bibnamefont {Genet}}, \bibinfo {author} {\bibfnamefont {J.~A.}\
  \bibnamefont {Hutchison}},\ and\ \bibinfo {author} {\bibfnamefont {T.~W.}\
  \bibnamefont {Ebbesen}},\ }\bibfield  {title} {\bibinfo {title} {Energy
  transfer between spatially separated entangled molecules},\ }\href
  {https://doi.org/10.1002/anie.201703539} {\bibfield  {journal} {\bibinfo
  {journal} {Ang. Chem. Int. Ed.}\ }\textbf {\bibinfo {volume} {56}},\ \bibinfo
  {pages} {9034} (\bibinfo {year} {2017})}\BibitemShut {NoStop}%
\bibitem [{\citenamefont {Welakuh}\ \emph {et~al.}(2021)\citenamefont
  {Welakuh}, \citenamefont {Ruggenthaler}, \citenamefont {Tchenkoue},
  \citenamefont {Appel},\ and\ \citenamefont {Rubio}}]{welakuh2021}%
  \BibitemOpen
  \bibfield  {author} {\bibinfo {author} {\bibfnamefont {D.~M.}\ \bibnamefont
  {Welakuh}}, \bibinfo {author} {\bibfnamefont {M.}~\bibnamefont
  {Ruggenthaler}}, \bibinfo {author} {\bibfnamefont {M.-L.~M.}\ \bibnamefont
  {Tchenkoue}}, \bibinfo {author} {\bibfnamefont {H.}~\bibnamefont {Appel}},\
  and\ \bibinfo {author} {\bibfnamefont {A.}~\bibnamefont {Rubio}},\ }\bibfield
   {title} {\bibinfo {title} {Down-conversion processes in ab-initio
  non-relativistic quantum electrodynamics},\ }\href
  {https://doi.org/10.1103/PhysRevResearch.3.033067} {\bibfield  {journal}
  {\bibinfo  {journal} {Phys. Rev. Research}\ }\textbf {\bibinfo {volume}
  {3}},\ \bibinfo {pages} {033067} (\bibinfo {year} {2021})}\BibitemShut
  {NoStop}%
\bibitem [{\citenamefont {Ruggenthaler}\ \emph {et~al.}(2018)\citenamefont
  {Ruggenthaler}, \citenamefont {Tancogne-Dejean}, \citenamefont {Flick},
  \citenamefont {Appel},\ and\ \citenamefont {Rubio}}]{ruggenthaler2017b}%
  \BibitemOpen
  \bibfield  {author} {\bibinfo {author} {\bibfnamefont {M.}~\bibnamefont
  {Ruggenthaler}}, \bibinfo {author} {\bibfnamefont {N.}~\bibnamefont
  {Tancogne-Dejean}}, \bibinfo {author} {\bibfnamefont {J.}~\bibnamefont
  {Flick}}, \bibinfo {author} {\bibfnamefont {H.}~\bibnamefont {Appel}},\ and\
  \bibinfo {author} {\bibfnamefont {A.}~\bibnamefont {Rubio}},\ }\bibfield
  {title} {\bibinfo {title} {From a quantum-electrodynamical
  light{\textendash}matter description to novel spectroscopies},\ }\href
  {https://doi.org/10.1038/s41570-018-0118} {\bibfield  {journal} {\bibinfo
  {journal} {Nature Reviews Chemistry}\ }\textbf {\bibinfo {volume} {2}},\
  \bibinfo {pages} {0118} (\bibinfo {year} {2018})}\BibitemShut {NoStop}%
\bibitem [{\citenamefont {Flick}\ \emph {et~al.}(2018)\citenamefont {Flick},
  \citenamefont {Rivera},\ and\ \citenamefont {Narang}}]{flick2018a}%
  \BibitemOpen
  \bibfield  {author} {\bibinfo {author} {\bibfnamefont {J.}~\bibnamefont
  {Flick}}, \bibinfo {author} {\bibfnamefont {N.}~\bibnamefont {Rivera}},\ and\
  \bibinfo {author} {\bibfnamefont {P.}~\bibnamefont {Narang}},\ }\bibfield
  {title} {\bibinfo {title} {Strong light-matter coupling in quantum chemistry
  and quantum photonics},\ }\href {https://doi.org/10.1515/nanoph-2018-0067}
  {\bibfield  {journal} {\bibinfo  {journal} {Nanophotonics}\ }\textbf
  {\bibinfo {volume} {7}},\ \bibinfo {pages} {1479} (\bibinfo {year}
  {2018})}\BibitemShut {NoStop}%
\bibitem [{\citenamefont {Wigner}(1932)}]{wigner1932}%
  \BibitemOpen
  \bibfield  {author} {\bibinfo {author} {\bibfnamefont {E.}~\bibnamefont
  {Wigner}},\ }\bibfield  {title} {\bibinfo {title} {On the quantum correction
  for thermodynamic equilibrium},\ }\href
  {https://doi.org/10.1103/PhysRev.40.749} {\bibfield  {journal} {\bibinfo
  {journal} {Phys. Rev.}\ }\textbf {\bibinfo {volume} {40}},\ \bibinfo {pages}
  {749} (\bibinfo {year} {1932})}\BibitemShut {NoStop}%
\bibitem [{\citenamefont {Walls}\ and\ \citenamefont
  {Milburn}(1994)}]{walls1994}%
  \BibitemOpen
  \bibfield  {author} {\bibinfo {author} {\bibfnamefont {D.}~\bibnamefont
  {Walls}}\ and\ \bibinfo {author} {\bibfnamefont {G.~J.}\ \bibnamefont
  {Milburn}},\ }\href {https://doi.org/10.1007/978-3-642-79504-6} {\emph
  {\bibinfo {title} {Quantum Optics}}}\ (\bibinfo  {publisher} {Springer-Verlag
  Berlin Heidelberg},\ \bibinfo {year} {1994})\BibitemShut {NoStop}%
\end{thebibliography}
\end{document}